\documentclass[a4paper,11pt]{article}
\pdfoutput=1 

\usepackage{jcappub} 

\usepackage[T1]{fontenc} 
\usepackage{color}

\usepackage{hyperref,breakurl}
\hypersetup{colorlinks=true, citecolor=blue, urlcolor=blue, linkcolor=blue}
\usepackage[usenames,dvipsnames]{xcolor}
\usepackage{dcolumn}
\usepackage{bm}
\usepackage{array}
\usepackage{graphicx}
\usepackage{graphics}
\usepackage{slashed}
\usepackage{subfig}
\usepackage{eso-pic}
\AddToShipoutPictureBG*{%
  \AtPageUpperLeft{%
    \hspace{0.75\paperwidth}%
    \raisebox{-3.5\baselineskip}{%
      \makebox[0pt][l]{\textnormal{DES-2022-0685}}
}}}%

\AddToShipoutPictureBG*{%
  \AtPageUpperLeft{%
    \hspace{0.75\paperwidth}%
    \raisebox{-4.5\baselineskip}{%
      \makebox[0pt][l]{\textnormal{FERMILAB-PUB-23-175-PPD}}
}}}%

\title{Cosmological constraints from the tomography of DES-Y3 galaxies with CMB lensing from ACT DR4 }

\author[1,2,3]{G.  A. Marques$^*$,} 
\emailAdd{gmarques@fnal.gov}

\affiliation[1]{Fermi National Accelerator Laboratory, P. O. Box 500, Batavia, IL 60510, USA}
\affiliation[2]{Department of Physics, Florida State University, Tallahassee FL, USA 32306}
\affiliation[3]{Kavli Institute for Cosmological Physics, University of Chicago, Chicago, IL 60637, USA}
\author[4,5]{M.~S.~Madhavacheril,}
\affiliation[4]{Department of Physics and Astronomy, University of Pennsylvania, Philadelphia, PA 19104, USA}
\affiliation[5]{Centre for the Universe, Perimeter Institute, Waterloo, ON N2L 2Y5, Canada}

\author[6]{O. Darwish,}
\affiliation[6]{Université de Genève, Département de Physique Théorique and Centre for Astroparticle Physics, 24 quai Ernest-Ansermet, CH-1211 Genève 4, Switzerland}

\author[7]{S. Shaikh,}
\affiliation[7]{School of Earth and Space Exploration, Arizona State University, Tempe, AZ, 85287, USA}

\author[8]{M.~Aguena,}
\affiliation[8]{Laborat\'orio Interinstitucional de e-Astronomia - LIneA, Rua Gal. Jos\'e Cristino 77, Rio de Janeiro, RJ - 20921-400, Brazil}

\author[9]{O.~Alves,}
\affiliation[9]{Department of Physics, University of Michigan, Ann Arbor, MI 48109, USA}

\author[10]{S.~Avila,}
\affiliation[10]{Institut de F\'{\i}sica d'Altes Energies (IFAE), The Barcelona Institute of Science and Technology, Campus UAB, 08193 Bellaterra (Barcelona) Spain}

\author[11]{D.~Bacon,}
\affiliation[11]{Institute of Cosmology and Gravitation, University of Portsmouth, Portsmouth, PO1 3FX, UK}

\author[12]{E.~J.~Baxter,}
\affiliation[12]{Institute for Astronomy, University of Hawai'i, 2680 Woodlawn Drive, Honolulu, HI 96822, USA}

\author[13]{K.~Bechtol,}
\affiliation[13]{Physics Department, 2320 Chamberlin Hall, University of Wisconsin-Madison, 1150 University Avenue Madison, WI  53706-1390}

\author[14]{M.~R.~Becker,}
\affiliation[14]{Argonne National Laboratory, 9700 South Cass Avenue, Lemont, IL 60439, USA}

\author[15,16]{E.~Bertin,}
\affiliation[15]{CNRS, UMR 7095, Institut d'Astrophysique de Paris, F-75014, Paris, France}
\affiliation[16]{Sorbonne Universit\'es, UPMC Univ Paris 06, UMR 7095, Institut d'Astrophysique de Paris, F-75014, Paris, France}

\author[58]{J.~Blazek,}

\author[66]{J. Richard Bond,}
\affiliation[66]{CITA, University of Toronto, Toronto ON M5S 3H8, Canada 
}

\author[17]{D.~Brooks,}
\affiliation[17]{Department of Physics \& Astronomy, University College London, Gower Street, London, WC1E 6BT, UK}

\author[18]{H. Cai,}
\affiliation[18]{Department of Physics and Astronomy, University of Pittsburgh, Pittsburgh, PA, USA 15260}

\author[37]{E. Calabrese,}
\affiliation[37]{School of Physics and Astronomy, Cardiff University, The Parade, Cardiff, Wales CF24 3AA, UK}

\author[19,8,20]{A.~Carnero~Rosell,}
\affiliation[19]{Instituto de Astrofisica de Canarias, E-38205 La Laguna, Tenerife, Spain}
\affiliation[20]{Universidad de La Laguna, Dpto. Astrofísica, E-38206 La Laguna, Tenerife, Spain}

\author[21,22]{M.~Carrasco~Kind,}
\affiliation[21]{Center for Astrophysical Surveys, National Center for Supercomputing Applications, 1205 West Clark St., Urbana, IL 61801, USA}
\affiliation[22]{Department of Astronomy, University of Illinois at Urbana-Champaign, 1002 W. Green Street, Urbana, IL 61801, USA}

\author[10]{J.~Carretero,}

\author[23]{R.~Cawthon,}
\affiliation[23]{Physics Department, William Jewell College, Liberty, MO, 64068}

\author[24,25]{M.~Crocce,}
\affiliation[24]{Institut d'Estudis Espacials de Catalunya (IEEC), 08034 Barcelona, Spain}
\affiliation[25]{Institute of Space Sciences (ICE, CSIC),  Campus UAB, Carrer de Can Magrans, s/n,  08193 Barcelona, Spain}

\author[8]{L.~N.~da Costa,}

\author[26]{M.~E.~S.~Pereira,}
\affiliation[26]{Hamburger Sternwarte, Universit\"{a}t Hamburg, Gojenbergsweg 112, 21029 Hamburg, Germany}

\author[27]{J.~De~Vicente,}
\affiliation[27]{Centro de Investigaciones Energ\'eticas, Medioambientales y Tecnol\'ogicas (CIEMAT), Madrid, Spain}

\author[28]{S.~Desai,}
\affiliation[28]{Department of Physics, IIT Hyderabad, Kandi, Telangana 502285, India}

\author[1]{H.~T.~Diehl,}
\author[17]{P.~Doel,}
\author[4,29]{C. Doux,}
\affiliation[29]{Universit\'e Grenoble Alpes, CNRS, LPSC-IN2P3, 38000 Grenoble, France}

\author[1,3,30]{A.~Drlica-Wagner,}
\affiliation[30]{Department of Astronomy and Astrophysics, University of Chicago, Chicago, IL 60637, USA}

\author[64,53]{J. Dunkley}

\author[31]{J.~Elvin-Poole,}
\affiliation[31]{Department of Physics and Astronomy, University of Waterloo, 200 University Ave W, Waterloo, ON N2L 3G1, Canada}

\author[32]{S.~Everett,}
\affiliation[32]{Jet Propulsion Laboratory, California Institute of Technology, 4800 Oak Grove Dr., Pasadena, CA 91109, USA}
\author[62,63]{Simone Ferraro,}
\affiliation[63]{Berkeley Center for Cosmological Physics, Department of Physics,
University of California, Berkeley, CA 94720, USA}

\author[33]{I.~Ferrero,}
\affiliation[33]{Institute of Theoretical Astrophysics, University of Oslo. P.O. Box 1029 Blindern, NO-0315 Oslo, Norway}

\author[1]{B.~Flaugher,}
 
\author[24,25]{P.~Fosalba,}
 
\author[34]{J.~Garc\'ia-Bellido,}
\affiliation[34]{Instituto de Fisica Teorica UAM/CSIC, Universidad Autonoma de Madrid, 28049 Madrid, Spain}

\author[4]{M.~Gatti,}
 
\author[10,3]{G.~Giannini,}

\author[35]{V. Gluscevic,}
\affiliation[35]{Department of Physics and Astronomy,
University of Southern California, Los Angeles, CA 90089, USA.}

\author[36]{D.~Gruen,}
\affiliation[36]{University Observatory, Faculty of Physics, Ludwig-Maximilians-Universit\"at, Scheinerstr. 1, 81679 Munich, Germany}

\author[21,22]{R.~A.~Gruendl,}
 
\author[1]{G.~Gutierrez,}

\author[37]{I.~Harrison,}

\author[69]{J.~Colin Hill,}
\affiliation[69]{Department of Physics, Columbia University, 538 West 120th Street, New York, NY, USA 10027}
\author[38]{S.~R.~Hinton,}
\affiliation[38]{School of Mathematics and Physics, University of Queensland,  Brisbane, QLD 4072, Australia}

\author[39]{D.~L.~Hollowood,}
\affiliation[39]{Santa Cruz Institute for Particle Physics, Santa Cruz, CA 95064, USA}

\author[40,41]{K.~Honscheid,}
\affiliation[40]{Center for Cosmology and Astro-Particle Physics, The Ohio State University, Columbus, OH 43210, USA}
\affiliation[41]{Department of Physics, The Ohio State University, Columbus, OH 43210, USA}

\author[9]{D.~Huterer,}

\author[17]{N.~Jeffrey,}

\author[4]{J. Kim,}

\author[43,44]{K.~Kuehn,}
\affiliation[43]{Australian Astronomical Optics, Macquarie University, North Ryde, NSW 2113, Australia}
\affiliation[44]{Lowell Observatory, 1400 Mars Hill Rd, Flagstaff, AZ 86001, USA}

\author[17]{O.~Lahav,}

\author[45,46]{P.~Lemos,}
\affiliation[45]{Department of Physics, Universit\'e de Montr\'eal, Montr\'eal, Canada}
\affiliation[46]{Ciela- Montreal Institute for Astrophysical Data Analysis and Machine Learning, Montréal, Canada}

\author[8]{M.~Lima,}
\affiliation{Departamento de F\'isica Matem\'atica, Instituto de F\'isica, Universidade de S\~ao Paulo, CP 66318, S\~ao Paulo, SP, 05314-970, Brazil}
 
\author[2]{K. M. Huffenberger,}

\author[47]{N.~MacCrann,}
\affiliation[47]{Department of Applied Mathematics and Theoretical Physics, University of Cambridge, Cambridge CB3 0WA, UK}
 
\author[48]{J.~L.~Marshall,}
\affiliation[48]{George P. and Cynthia Woods Mitchell Institute for Fundamental Physics and Astronomy, and Department of Physics and Astronomy, Texas A\&M University, College Station, TX 77843,  USA}

\author[27]{J. Mena-Fern{\'a}ndez,}

\author[49,10]{R.~Miquel,}
\affiliation[49]{Instituci\'o Catalana de Recerca i Estudis Avan\c{c}ats, E-08010 Barcelona, Spain}

\author[36,50]{J.~J.~Mohr,}
\affiliation[50]{Max Planck Institute for Extraterrestrial Physics, Giessenbachstrasse, 85748 Garching, Germany}

\author[71,72]{K. Moodley,}
\affiliation[71]{A. Astrophysics Research Centre, University of KwaZulu-Natal, Westville Campus, Durban 4041, South Africa}
\affiliation[72]{B. School of Mathematics, Statistics \& Computer Science, University of KwaZulu-Natal, Westville Campus, Durban 4041, South Africa}

\author[51]{J.~Muir,}
\affiliation[51]{Perimeter Institute for Theoretical Physics, 31 Caroline St. North, Waterloo, ON N2L 2Y5, Canada}

\author[33]{S. Naess,}

\author[67]{F. Nati,}
\affiliation[67]{Department of Physics, University of Milano-Bicocca, Piazza della Scienza 3, 20126 Milano (MI), Italy}

\author[64]{L. A. Page,}
\affiliation[64]{Joseph Henry Laboratories of Physics, Jadwin Hall, Princeton University, Princeton, NJ, USA 08544}

\author[52]{A.~Palmese,}
\affiliation[52]{Department of Physics, Carnegie Mellon University, Pittsburgh, Pennsylvania 15312, USA}

\author[53]{A.~A.~Plazas~Malag\'on,}
\affiliation[53]{Department of Astrophysical Sciences, Princeton University, Peyton Hall, Princeton, NJ 08544, USA}

\author[40,41]{A.~Porredon,}
\affiliation{Institute for Astronomy, University of Edinburgh, Edinburgh EH9 3HJ, UK}

\author[30, 3]{J.~Prat,}

\author[57,65]{F. J. Qu,}
\affiliation[65]{DAMTP, Centre for Mathematical Sciences, Wilberforce Road, Cambridge CB3 0WA, UK}

\author[54]{M.~Raveri,}
\affiliation[54]{Department of Physics, University of Genova and INFN, Via Dodecaneso 33, 16146, Genova, Italy}

\author[40]{A.~J.~Ross,}

\author[55,56]{E.~S.~Rykoff,}
\affiliation[55]{Kavli Institute for Particle Astrophysics \& Cosmology, P. O. Box 2450, Stanford University, Stanford, CA 94305, USA}
\affiliation{[56]SLAC National Accelerator Laboratory, Menlo Park, CA 94025, USA}

\author[47,57]{G. S. Farren,}
\affiliation[57]{Kavli Institute for Cosmology Cambridge, Madingley Road, Cambridge CB3 0HA, UK}

\author[58]{S.~Samuroff,}
\affiliation[58]{Department of Physics, Northeastern University, Boston, MA 02115, USA}

\author[27]{E.~Sanchez,}

\author[9]{M.~Schubnell,}

\author[73]{N. Sehgal,}
\affiliation[73]{Physics and Astronomy Department, Stony Brook University, Stony Brook, NY USA 11794}

\author[27]{I.~Sevilla-Noarbe,}

\author[59]{E.~Sheldon,}
\affiliation[59]{Brookhaven National Laboratory, Bldg 510, Upton, NY 11973, USA}

\author[65,57]{B. D. Sherwin,}

\author[70]{C. Sif\'on,}
\affiliation[70]{Instituto de F\'isica, Pontificia Universidad Cat\'olica de Valpara\'iso, Casilla 4059, Valpara\'iso, Chile}

\author[60]{M.~Smith,}
\affiliation[60]{School of Physics and Astronomy, University of Southampton,  Southampton, SO17 1BJ, UK}

\author[63,53]{D. N. Spergel,}
\affiliation[63]{Center for Computational Astrophysics, Flatiron Institute, 162 5th Avenue, New York, NY 10010 USA}

\author[64]{S. T. Staggs,}

\author[61]{E.~Suchyta,}
\affiliation[61]{Computer Science and Mathematics Division, Oak Ridge National Laboratory, Oak Ridge, TN 37831}

\author[9]{G.~Tarle,}

\author[40]{C.~To,}

\author[7]{A. Van Engelen,}
 
\author[9,62]{N.~Weaverdyck,}
\affiliation[62]{Lawrence Berkeley National Laboratory, 1 Cyclotron Road, Berkeley, CA 94720, USA}

\author[50,36]{J.~Weller,}

\author[68]{L. Wenzl,}
\affiliation[68]{Department of Astronomy, Cornell University, Ithaca, NY, 14853, USA}
\author[60]{P.~Wiseman,}

\author[42]{E. J. Wollack,}
\affiliation[42]{NASA Goddard Space Flight Center, Greenbelt, MD 20771, USA}

\author[1]{B.~Yanny}

\collaboration{ACT \& DES Collaborations}
 
\date{\today}

\abstract{We present a measurement of the cross-correlation between the \maglim galaxies selected from the Dark Energy Survey (DES) first three years of observations (Y3) and cosmic microwave background (CMB) lensing from the Atacama Cosmology Telescope (ACT) Data Release 4 (DR4), reconstructed over $\sim 436$ $\sqdeg$ of the sky. Our galaxy sample, which covers $\sim 4143$ $\sqdeg$, is divided into six redshift bins spanning the redshift range of $0.20<z<1.05$. We adopt a blinding procedure until passing all consistency and systematics tests. After imposing scale cuts for the cross-power spectrum measurement, we reject the null hypothesis of no correlation at 9.1$\sigma$. We constrain cosmological parameters from a joint analysis of galaxy and CMB lensing-galaxy power spectra considering a flat \LCDM model, marginalized over 23 astrophysical and systematic nuisance parameters. We find the clustering amplitude $S_8\equiv \sigma_8 (\Omega_m/0.3)^{0.5} = 0.75^{+0.04}_{-0.05}$. In addition, we constrain the linear growth of cosmic structure as a function of redshift. Our results are consistent with recent DES Y3 analyses and suggest a preference for a lower $S_8$ compared to results from measurements of CMB anisotropies by the \textit{Planck} satellite, although at a mild level ($< 2 \sigma$) of statistical significance. }

\keywords{Cosmology, Large-scale structure, CMB lensing cross-correlations, Growth of Cosmic Structure, Surveys}

\begin{document}
\maketitle
\flushbottom

\section{Introduction}
According to the current favored cosmological model, most of the energy density of the Universe is composed of cold dark matter (\rm{CDM}) and a cosmological constant $\Lambda$. The tightest constraints on the $\Lambda$\rm{CDM} cosmological parameters come from the measurements of the Cosmic Microwave Background (CMB) \citep{planckparams}, which provide a snapshot of the very early universe. However, the CMB data also offers a great opportunity to constrain information from the later evolution of the Universe, for example, by studying the weak gravitational lensing (WL) of the CMB photons due to the gravitational potentials of large-scale structure (LSS) \citep{lewis2006weak}. 

Over the past two decades, observations from large galaxy imaging surveys have become one of the key ways of probing cosmology. The early results of Stage III photometric galaxy surveys (as defined by the Dark Energy Task Force report \citep{albrecht2006report}), such as the Dark Energy Survey (DES) \citep{flaugher2015dark,abbott2019cosmological,porredon2021dark_maglim,abbott2021dark}, Hyper Suprime-Cam Subaru Strategic Program (HSC-SSP) \citep{hikage2019cosmology,hamana2020cosmological} and Kilo-Degree Survey (KiDS) \citep{joudaki2018kids,van2018kids+} have demonstrated the feasibility of cataloging millions of galaxies, constraining the geometry and structure growth of the Universe, and testing for systematic effects. Despite the remarkable success in tightening the cosmological parameter constraints and testing the $\Lambda$\rm{CDM} model, the results of these LSS observations reveal a possible tension in the derived parameter $S_{8}= \sigma_{8}\sqrt{\Omega_{m}/0.3}$, where $\sigma_{8}$ is the amplitude of mass fluctuations parametrized as the standard deviation of the linear overdensity fluctuations in 8 $h^{-1}$\rm{Mpc} spheres at the present time, $h$ is the dimensionless Hubble parameter, and $\Omega_{m}$ is the density parameter of matter at the present time. In particular, the value of $S_{8}$ obtained from \textit{Planck} CMB TT+TE+EE+lowE data ($z = 1100)$ assuming $\Lambda$\texttt{CDM} is $S_{8} = 0.834 \pm 0.016$ \citep{planckparams}, while cosmic shear analyses ($z \lesssim 1)$ report a $1.7\sigma- 3.4\sigma$ lower value, e.g.\cite{heymans2013cfhtlens,abbott2018dark,hikage2019cosmology,troster2020cosmology, heymans2021kids,asgari2021kids,survey2023y3+}. In addition, CMB lensing analyses ($z \sim 0.5-5$) from \textit{Planck} \citep{planckcmblens}, South Pole Telescope (SPT) \citep{spt_lensing2023} and ACT \citep{dr6atacama, qu2023atacama}, reveal an $S_8$ consistent with those inferred using early Universe CMB data,  favoring higher values compared to those inferred from shear and galaxy clustering measurements at lower redshifts and smaller scales.

The upcoming generation of LSS surveys, including the Rubin Observatory Legacy Survey of Space and Time (LSST) \citep{ivezic2019lsst}, DESI \citep{aghamousa2016desi}, \texttt{SPHEREx} \citep{dore2014cosmology}, Euclid \citep{laureijs2011euclid}, Nancy Grace Roman Space Telescope (Roman) \citep{spergel2015wide}, as well as CMB experiments like SPT-3G \citep{benson2014spt}, Simons Observatory (SO) \citep{ade2019simons} and CMB-S4 \citep{abazajian2019cmb}, hold the potential to shed light on the source of the possible discrepancy between early and late-time probes of the Universe. In addition, innovative data analysis methods can help overcome systematic effects and shed light on the physical properties of the Universe.\

The combination of distinct cosmological probes has been shown to be a promising technique for investigating the properties of our Universe on cosmic scales. In particular, cross-correlations of CMB lensing with galaxy surveys offer an opportunity to probe cosmological parameters while having more stringent control over additive observational systematic effects. Examples of this approach include the cross-correlation of CMB lensing with galaxy weak lensing \cite{hand2015first,liu2015cross,nicola2016integrated,omori2019dark_wl,namikawa2019evidence,marques2020cross,robertson2021strong,DESXSPT}, galaxy density \citep{bianchini2015cross,bianchini2018constraining,omori2019dark,krolewski2020unwise,marques2020tomographic,darwish2021atacama,kitanidis2021cross,hang2021galaxy,garcia2021growth,krolewski2021cosmological,white2022cosmological}, quasar density \citep{sherwin2012atacama,geach2013direct,doux2018cosmological}, cosmic infrared background \citep{holder2013cmb,van2015atacama,cao2020cross}, and many others.

In this work, we measure the cross-correlation between the CMB lensing from  the fourth data-release (DR4) of the Atacama Cosmology Telescope (ACT) \citep{darwish2021atacama, aiola2020atacama} and the galaxy density from the magnitude-limited sample (\maglim) of the DES Year-3 data release (Y3) \citep{porredon2021dark_maglim}. We consider six redshift bins spanning the redshift range of $0.20<z<1.05$, over $\sim 436$ square deg. of the sky. We jointly model the galaxy and the galaxy-CMB lensing power spectra to probe the amplitude and evolution of linear galaxy bias, linear growth of
structure, and cosmological parameters such as $\Omega_m$ and $S_8$.
 
This paper is structured as follows: in Sec. \ref{sec:th} we describe the theoretical framework used to model the galaxy and galaxy-CMB lensing power spectra. In Sec. \ref{sec:data} we present the data sets used in our analysis. In Sec. \ref{sec:methods}, we describe the method to estimate the angular power spectrum, covariance matrix, and parameter inference. We validate our methodology using simulations in Sec. \ref{sec:test_logmocks} and we study the impact of systematic effects in Sec. \ref{sec:sys}. The main results are discussed in Sec. \ref{sec:results}. Finally, in Sec. \ref{sec:conclusions} we present our conclusions.

\section{Theory}
\label{sec:th}

The CMB lensing convergence $\kappa^{\rm CMB}$ is expected to be correlated with the galaxy overdensity $\delta_{\rm g}$, since both are tracers of the underlying mass distribution. On linear scales, these observables are the weighted projection of 
the three-dimensional matter density contrast $\delta$ along the line-of-sight
\begin{equation}
    u(\mathbf{\hat{n}}) = \int_{0}^{\infty}dz  ~q^{u}(z) \delta(\chi(z)\mathbf{\hat{n}},z),
\end{equation}
where the fields $u$ = $\{\kappa^{\rm CMB}, \delta_{\rm g} \}$ are
defined on the celestial sphere, and $q^{u}(z)$ are their respective projection kernels. For a flat universe, the CMB lensing kernel is given by
\begin{equation}
    q^{\kappa^{\rm CMB}}(z)= \frac{3 \Omega_{m}}{2c}\frac{H_{0}^2}{H(z)}(1+z) \chi(z)\frac{\chi_{*}-\chi(z)}{\chi_{*}}.
\label{eq:cmb_kernel}
\end{equation}
Here, $c$ is the speed of light, $H(z)$ is the Hubble parameter at redshift $z$, and $H_{0}$ is the present-day value of the Hubble parameter. The terms $\chi(z)$ and $\chi_{*}$ denote the comoving distance to redshift $z$ and to the last-scattering surface, respectively.

The galaxy kernel $q^{\delta_{g}}$ is expressed as the sum of two terms, 
 \begin{equation}
     q^{\delta_{\rm g,i}}(z) = b^{i}(z)n_{\rm g}^{i}(z) + \mu^{i}(z),
\label{eq:galaxy_kernel}
 \end{equation}
where $i$ labels the redshift bin, $n_{g}$ is the normalized redshift distribution of the galaxy sample, and $b(z)$ is the galaxy bias. We model $b(z)$ of our sample assuming a linear, deterministic, and scale-independent galaxy bias \citep{fry1993biasing} that is constant across each tomographic bin, as validated in \cite{krause2021dark}. The second term, $\mu$, quantifies the so-called magnification bias effect, caused by the increase of the observed flux of the background galaxies due to lensing by foreground structure. This effect leads to a deviation in the observed number density of galaxies as it allows the detection of galaxies that, in the absence of lensing, would be fainter than the limiting magnitude of the survey \citep{turner1980effect,schneider1989number,moessner1998angular}. The magnification bias correction is given by
\begin{equation}
    \mu^{i}(z) = (5s^{i} -2)\frac{3}{2c} \frac{\Omega_{m}H_{0}^2}{H(z)}(1+z)\chi(z)
    \int_{z}^{z^{*}} dz' n_{g}^{i}(z') \frac{(\chi(z')-\chi(z))}{\chi(z')}. 
\label{eq:mag_relation2kappa}
\end{equation}
The quantity $s$ denotes the slope of the  cumulative apparent magnitude distribution  $N^{\rm int}(m)$
\begin{equation}
    s^{i} = \frac{d\log N^{\rm int,i}(m)}{d m}.
\end{equation}

The values of $s$ for the DES Y3 galaxy sample are estimated in \cite{y3-2x2ptmagnification} and are held fixed to the values listed in Table \ref{tab:params}. More discussion on the magnification estimation and the robustness of the assumption of constant values within the tomographic bins can be found in \cite{krause2021dark, derose2021dark,y3-2x2ptmagnification}.  

Following previous DES-Y3 analyses \citep{descollaboration2021dark, porredon2021dark_maglim}, we model the uncertainty in the source galaxy redshift distributions with an additive shift parameter, $\Delta z^{i}$, where $i$ labels the redshift bin. The photometric redshift distribution is modified as
\begin{equation}
    n_{g}^{i}(z) \rightarrow n_{g}^{i} (z- \Delta z^{i}).
\end{equation}
We also parametrize the uncertainty on the width of the redshift distribution by a stretch parameter $\sigma_{z}^{i}$ \citep{cawthon2020dark, porredon2021dark_maglim}, so that the combination with the shift parameter leads to
\begin{equation}
    n_{g}^{i}(z) \rightarrow \sigma_{z}^{i}n_{g}^{i}(\sigma_{z}^{i}[z - \langle z \rangle ] +\langle z \rangle - \Delta z^{i}).
\end{equation}

Assuming the Limber approximation \citep{limber1953analysis}, the theoretical galaxy-galaxy and galaxy-CMB lensing angular power spectra of a given $i$-redshift bin can be evaluated as 
\begin{equation}
\begin{aligned}
C_{\ell}^{gg, ii} &= \int_{0}^{\infty}\frac{dz}{c}\frac{H(z)}{\chi^2(z)}[q^{\delta_{g,i}}(z)]^2 P\bigg(k=\frac{\ell+\frac{1}{2}}{\chi(z)},z \bigg), \\
C_{\ell}^{\kappa g,i} &= \int_{0}^{\infty}\frac{dz}{c}\frac{H(z)}{\chi^2(z)}q^{\kappa^{\rm CMB}}(z)q^{\delta_{g,i}}(z) P\bigg(k=\frac{\ell+\frac{1}{2}}{\chi(z)},z \bigg),
\end{aligned}
\label{eq:powerspe_th}
\end{equation}
where $P(k,z)$ is the nonlinear matter power spectrum. We model $P(k,z)$ using the \texttt{CAMB} Boltzmann code \citep{lewis2000efficient} with the \texttt{HALOFIT} prescription of \cite{takahashi2012revising} and we use the DESC Core Cosmology Library (\texttt{CCL)} \citep{chisari2019core} to compute the theoretical quantities.  

For a $\Lambda$\rm{CDM} Universe, in which the only relevant density contrast is that of pressureless matter, the matter power spectrum in the Eq. \ref{eq:powerspe_th} can be written in terms of the normalized linear growth function $D(z)$ as
\begin{equation}
    P(k,z) = P(k,0)D^2(z).
\end{equation} 
From Eqs. \ref{eq:cmb_kernel}, \ref{eq:galaxy_kernel}, and \ref{eq:powerspe_th} it is possible to notice that $C_{\ell}^{\kappa g}$ is sensitive to $b D^2(z)$, while $C_{\ell}^{gg}$ is sensitive to $b^2 D^2(z)$. The combination of these two observed (obs) quantities allows us to break the degeneracy between the linear galaxy bias and linear growth through the $\hat{D}_{G}$ estimator \citep{giannantonio2016cmb}, defined as 
\begin{equation}
\hat{D}_{G}^{i} \equiv \bigg\langle \frac{(C_{\ell}^{\kappa g,i})_{\rm obs}}{(\slashed{C}_{\ell}^{\kappa g,i})_{\rm th}}\sqrt{\frac{(\slashed{C}_{\ell}^{gg,ii})_{\rm th}}{(C_{\ell}^{gg,ii})_{\rm obs}}} \bigg\rangle_{\ell}.
\label{eq:dg}
\end{equation}

In the above equation, the brackets denote an average over the range of multipoles included in the analysis, and $\slashed{C}_{\ell}^{gg}$ and $\slashed{C}_{\ell}^{\kappa g}$ are the theoretical power spectra defined in Eq. \ref{eq:powerspe_th} evaluated with the matter power spectrum at $z=0$, 
\begin{equation}
\begin{aligned}
\slashed{C}_{\ell}^{gg,ii} &= \int_{0}^{\infty}\frac{dz}{c}\frac{H(z)}{\chi^2(z)}[q^{\delta_{g,i}}(z)]^2 P(k, z=0), \\
\slashed{C}_{\ell}^{\kappa g,i} &= \int_{0}^{\infty}\frac{dz}{c}\frac{H(z)}{\chi^2(z)}q^{\kappa^{\rm CMB}}(z)q^{\delta_{g,i}}(z) P(k, z=0).
\end{aligned}
\end{equation}
These terms are introduced to keep $\hat{D}_{G}$ bias-independent and normalized to unity today, i.e., $\hat{D}_{G}(z=0) = 1$.  

Given the current discrepancies in the literature concerning the values of $S_8$, it is crucial to emphasize that the aforementioned formalism is based on a set of assumptions. If stochasticity is present due to, for example, physical processes impacting the halo collapse or any systematics in the galaxy selection, the galaxy bias in the $C_{\ell}^{gg}$ term would absorb a stochastic component in a distinct manner compared to the $C_{\ell}^{\kappa g}$ term, thereby affecting the $S_8$ amplitude. Further understanding of potential systematic issues associated with stochasticity is an important next step, especially for future high-signal-to-noise measurements.

\section{DATA}
\label{sec:data}
\subsection{ACT CMB lensing}
\label{sec:cmblen_data}
The CMB lensing convergence map used in this work \citep{darwish2021atacama} is reconstructed from the CMB temperature and polarization data of the fourth ACT data release (DR4) \footnote{The ACT CMB lensing DR4 products are publicly available at \url{https://lambda.gsfc.nasa.gov/product/act/actpol_prod_table.cfm}} \citep{thornton2016atacama, aiola2020atacama, choi2020atacama}. The CMB data used to perform the lensing reconstruction were synthesized from observations taken during the 2014 and 2015 seasons in the 98 GHz and 150 GHz frequency bands in two different patches of the sky, namely BOSS-North (\texttt{BN}) and deep-56 (\texttt{d56}). In particular, we use CMB lensing reconstructed in the \texttt{d56} region, which overlaps with DES Y3 data over a total area of $\approx 456$ square degrees of the sky. This is $\sim 11\%$ of the overlapping region of the most recent ACT lensing map from DR6 \citep{dr6atacama}, which will be explored in future work.

The CMB lensing map is reconstructed by applying a quadratic estimator \citep{hu2002mass}, with a minimum variance combination of temperature and polarization CMB data. Modes with $\ell < 500 $ and $\ell > 3000 $ are removed to restrict to scales where the ACT map-maker transfer function is nearly unity and to minimize foreground contamination, respectively. 

One of the largest potential contaminants affecting cross-correlation analysis with CMB lensing comes from the thermal Sunyaev-Zel’dovich effect (tSZ) \citep{van2014cmb, schaan2019foreground,madhavacheril2020atacama}. The ACT DR4 lensing maps come in two versions: The baseline map, which combines ACT and multi-frequency data from \textit{Planck} to reconstruct CMB lensing using an extended multipole range from $100< \ell< 3350$, and a second version that only uses ACT data from the 2 frequency bands. The baseline map has the tSZ effect deprojected using the method described in \cite{madhavacheril2018mitigating,darwish2021atacama}.

We use the baseline ACT+\textit{Planck} tSZ-free map in our main analysis. However, in Sec. \ref{sec:sys} we investigate the impact of the tSZ in our results by repeating our analysis using the lensing map without the tSZ mitigation. During the lensing reconstruction, an apodized mask is applied to the input CMB maps \citep{darwish2021atacama}, and the released lensing maps are divided by the mean of the square of the mask, $\langle W_{c}^2 \rangle$, in order to correct for the loss of power resulting from the original masking. In our analysis, we multiply the CMB lensing map by $\langle  W_{c}^2\rangle$ because we consider the mask correction in the power spectrum computation. 

We convert the original CMB lensing maps and the associated mask to \texttt{HEALPix} \citep{zonca2019healpy} format with resolution $N_{\rm side}= 1024$ using the pixell\footnote{\url{https://github.com/simonsobs/pixell}} package. In particular, we use the \\ \texttt{healpix\_from\_enmap}~ routine with $\ell_{\rm max}$ of 6000 to project the maps. The total effective area in our cross-correlation study is equal to $\sim 436$ \rm{deg}$^2$.

\begin{figure}
   \begin{center}
    \includegraphics[width=\linewidth]{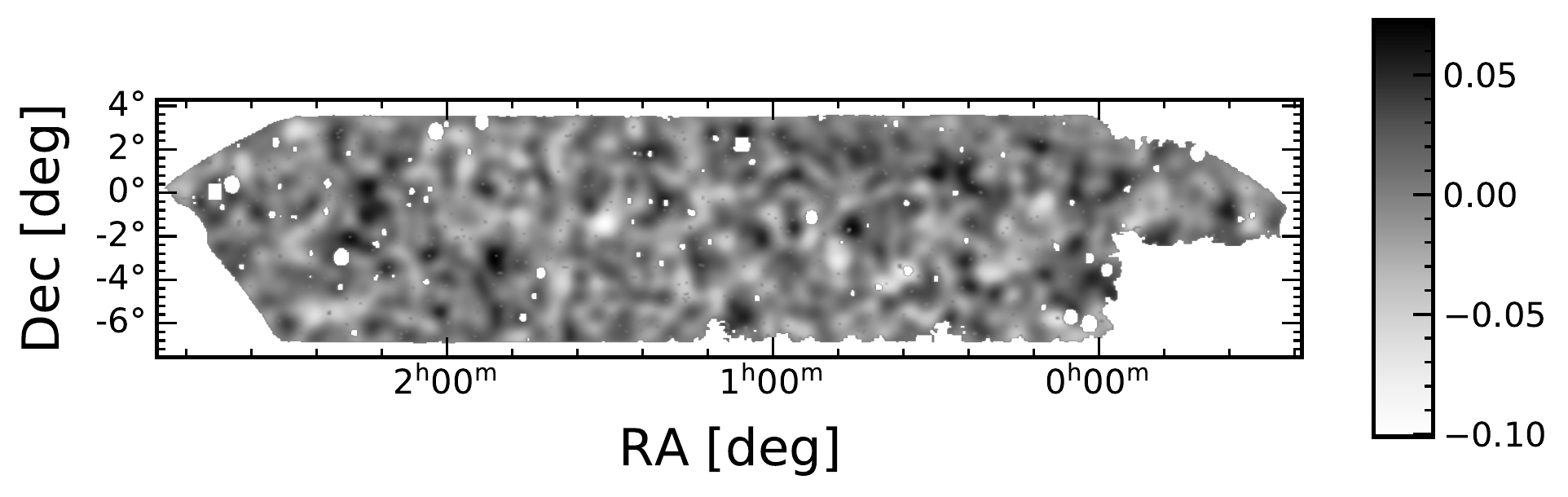}
     \caption{CMB convergence map of the \texttt{d56} region, reconstructed from the combination of tSZ-cleaned ACT and \textit{Planck} data \citep{darwish2021atacama}. The map is smoothed with a Gaussian beam with 20 arcmin for visualization purposes only. Data in the white regions are masked out regions. }
    \label{fig:act}
\end{center}
\end{figure}

The estimation of the ACT convergence map and its associated products are described in more detail in \cite{darwish2021atacama}. The reconstructed ACT CMB lensing map is shown in Figure \ref{fig:act}. In this figure, the CMB lensing map is smoothed with a Gaussian kernel on a scale of 20 arcmin for visualization purposes only.

\subsection{DES Magnitude-limited sample (\maglim)}
\label{sec:des_data}
The DES Y3 data relies on observations taken from August 2013 to February 2016, in five broad filters, $grizY$, using the Dark Energy Camera \citep{Flaugher:2015}. The main catalog, referred to as \texttt{Y3 GOLD}, includes nearly 400 million objects over $\sim 5000 \deg^2$ of the sky, with depth reaching S/N $\sim10$ up to limiting magnitudes of $g=24.3$, $r=24.0$, $i=23.3$, $z=22.6$, and $Y=21.4$ \citep{y3-gold}. 

\begin{figure}[h!]
 \begin{center}
    \includegraphics[scale=0.6]{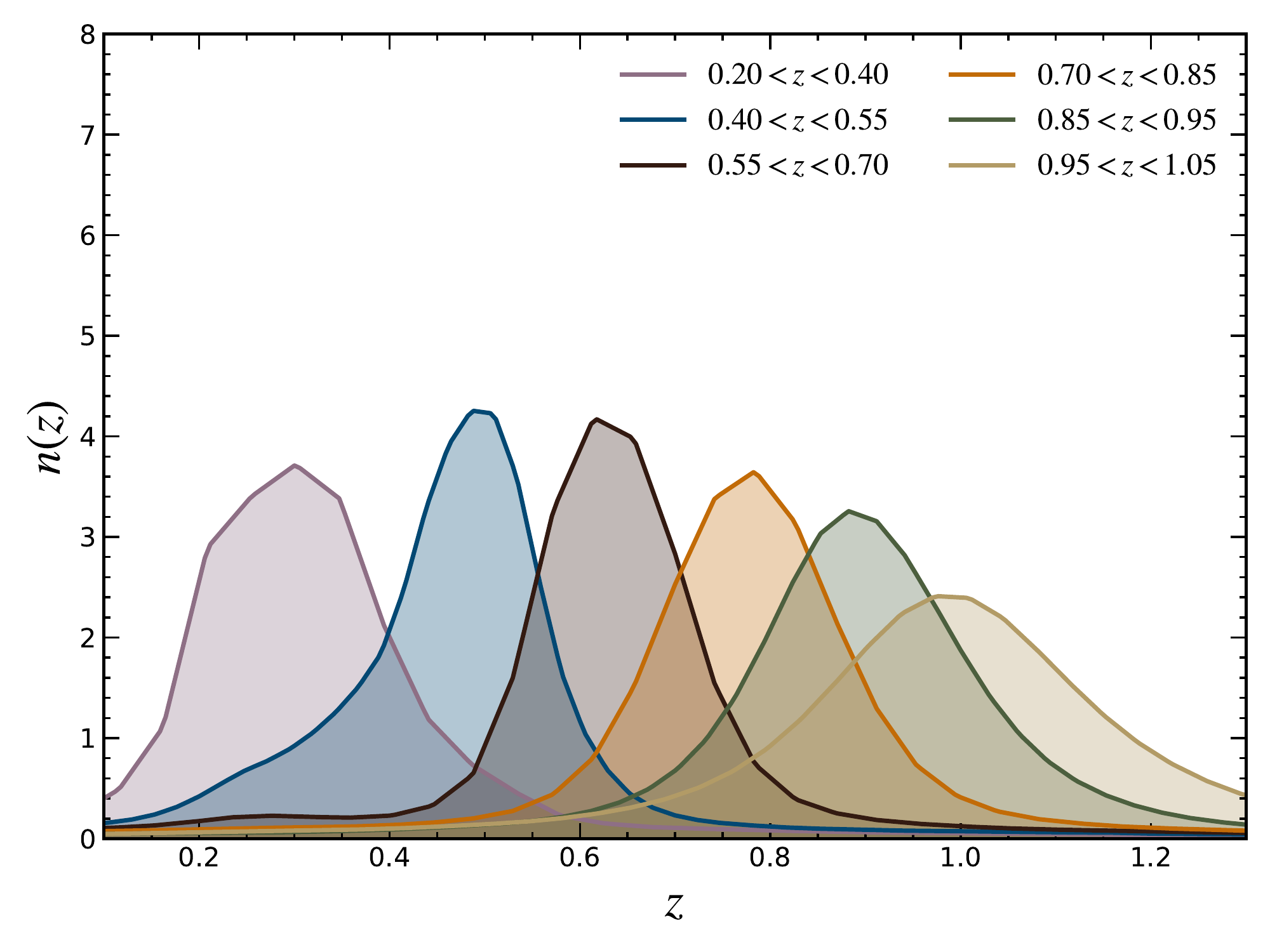}
    \caption{Normalized redshift distributions of the DES Y3 \maglim sample.}
    \label{fig:nz}
    \end{center}
\end{figure}

In this work, we use the \maglim catalog, which is a sample defined with a magnitude cut in the $i$-band that depends linearly on photometric redshift: $i< 4 z +18$ \citep{porredon2021dark}. This selection is based on an optimization found in \cite{porredon2021dark}.
Additionally, selecting $i> 17.5$ removes residual stellar contamination  and other bright objects. The photometric redshift is estimated from the Directional Neighborhood Fitting (DNF) algorithm \citep{dnfpaper,y3-gold} and has been validated using cross-correlations with spectroscopic galaxies \citep{cawthon2020dark}. We split the catalog into 6 tomographic photometric redshift bins, spanning the range of $0.20<z<1.05$, with bin edges [0.20, 0.40, 0.55, 0.70, 0.85, 0.95, and 1.05]. In Fig. \ref{fig:nz} we show the normalized redshift distribution for each tomographic bin.

In \cite{rodriguez2021dark} it was found that the DES galaxy number density fluctuates as a function of a series of observational conditions and survey properties. To correct this dependency in the number counts, we weighted each galaxy by the inverse of the angular selection function at that galaxy's location \citep{porredon2021dark_maglim,descollaboration2021dark}. More information about the \maglim systematic tests and the methodology to assign weights can be found in \cite{rodriguez2021dark} and the impact of these systematics is also investigated in Sec. \ref{sec:sys}.

\subsubsection{Pixelized masks and maps}
\label{sec:masks_maps}
The galaxies in the \maglim catalog have an associated weight to correct for the impact of observational systematics \cite{rodriguez2021dark}. We use these weights to construct a map of the number of sources in each pixel $\rm p$, estimated as $N_{\rm p} = \sum_{i\in \rm p}w_{i}$, where $w_{i}$ is the weight of the $i$-th galaxy. We create pixelized maps of the galaxy overdensity at a HEALPix resolution parameter $N_{\rm side}
=1024$, defined by
\begin{equation}
    \delta_{\rm p} = \bigg(\frac{1}{f_{\rm p}} \frac{N_{\rm p}}{\bar{n}} \bigg) -1,
\end{equation}
where $\bar{n}$ denotes the mean of all number of sources in the unmasked pixels computed as
\begin{equation}
    \bar{n}  = \frac{\sum_{\rm p}N_{\rm p}}{(\sum_{\rm p}f_{\rm p})}.
\end{equation}
The term $f_{\rm p}$ represents the DES mask, which contains information about the fraction that each pixel has been observed, the so-called fractional coverage or completeness. This mask is constructed by a series of cuts to remove astrophysical foregrounds or regions with recognized systematics, e.g., bright stars or
regions with data processing issues \citep{porredon2021dark_maglim}. Specifically,
regions with bad quality, extreme observing
conditions, areas outside the survey footprint, and pixels with fractional coverage smaller
than $80\%$ are removed. More details about these
cuts and the flags imposed on the \maglim sample can be found in
\citep{sevilla2021dark,porredon2021dark_maglim}.

In its original resolution of $N_{\rm side} = 4096$, the $f_{\rm p}$ map has values ranging from $0.8$ to $1.0$ within the DES footprint, and $0.0$ outside the footprint. We degrade the completeness map from the original resolution to $N_{\rm side} =1024$ by setting the $f_{\rm p}$ values as the mean of the higher resolution pixels. After proper masking, the \maglim sample contains $\sim 10$ million galaxies covering an area of $\sim 4100 \deg^2$ of the sky and $\sim 1$ million galaxies in the ACT \texttt{d56} region. Some of the key features of the galaxy sample are summarized in Table \ref{table:specifics}.



\begin{table}
\centering
\begin{tabular}{|l|l|l|l|l|}
\hline
Redshift range     & $n_{\rm gal}$ {[}arcmin$^{-2}${]} & $\ell_{\rm max}^{gg}$ & $\ell_{\rm max}^{\kappa g}$ \\ \hline
\
$0.20<z<0.40$             &  0.150                              & 97 &  137                                         \\
$0.40<z<0.55$            &  0.107                              & 144 & 175                                             \\
$0.55<z<0.70$           &  0.109                              & 187 & 227                                             \\
$0.70<z<0.85$            &  0.146                              & 223 & 271                                           \\
$0.85<z<0.95$             &  0.106                              & 248 & 301                                             \\
$0.95<z<1.05$            &  0.100                              & 268 & 325                                             \\ \hline
\end{tabular}
\caption{Summary of properties of the \maglim lens sample: Redshift range, the effective number density of galaxies in units of
\rm{arcmin}$^{-2}$, and maximum multipole considered in the cosmological
analysis for the $C_{\ell}^{gg}$ and $C_{\ell}^{\kappa g}$  (Sec \ref{sec:method_PS}).
}
\label{table:specifics}
\end{table}

\begin{figure}[ht]
    \begin{center}
    \includegraphics[width=\linewidth]{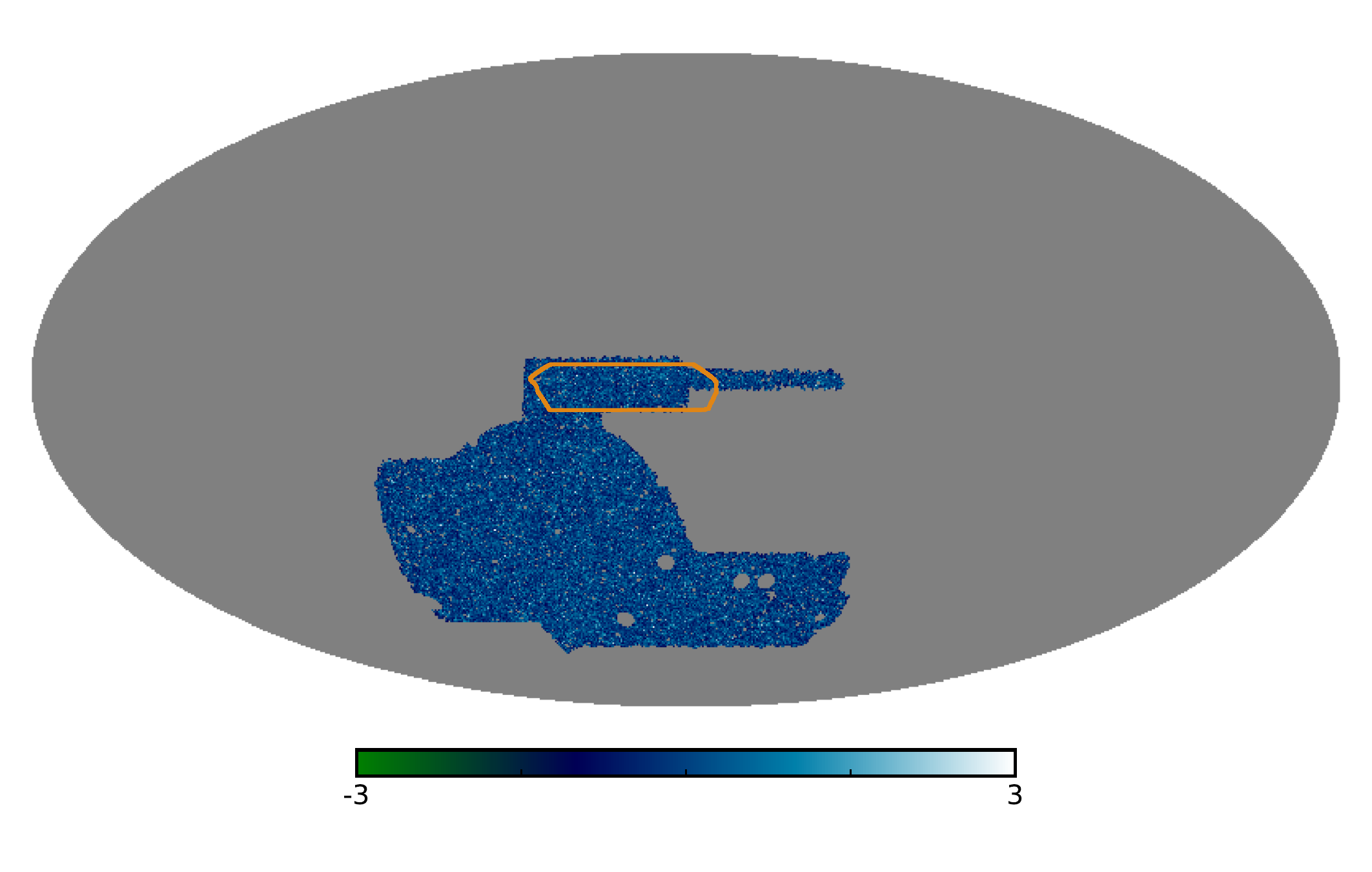}
     \caption{ \maglim overdensity map for the full sample in Mollweide projection and equatorial coordinates. The grey region shows parts of the sky that are masked, and the orange line indicates the outline of the ACT \texttt{d56} footprint.}
    \label{fig:dlt_des}
    \end{center}
\end{figure}

Fig. \ref{fig:dlt_des} shows the \maglim overdensity map in the redshift range of $0.20<z<1.05$. The grey region shows the region of the sky that is masked, and the orange line represents the ACT \texttt{d56} footprint.  

\section{METHODS}
\label{sec:methods}
\subsection{Angular power spectrum }
\label{sec:method_PS}

The angular power spectrum estimate for a survey that covers only a fraction of the sky is affected by the mask, which introduces coupling between the different modes of the true power spectrum. We use a pseudo-$C_{\ell}$ estimator \citep{hivon2002master} implemented by \texttt{NaMaster}~\citep{alonso2019unified} to obtain an unbiased estimate of the angular power spectra. Roughly speaking, the true underlying power spectrum $C_{\ell}^{\rm true}$ is recovered in the pseudo-$C_{\ell}$ approach at every integer $\ell$ by inverting 
\begin{equation}
    C_{\ell}^{\rm obs} = \sum_{\ell'} M_{\ell\ell'} C_{\ell'}^{\rm true},
    \label{eq:pseudo-cl}
\end{equation}
where the $C_{\ell}^{\rm obs}$ is the observed power spectrum and $M$ is the so-called mode-coupling matrix. It is worth noticing that the mode coupling matrix depends solely on the mask information.\ 

We compute the galaxy-power spectrum considering the full DES footprint, as defined by the DES mask described in Sec.\ref{sec:masks_maps}. When computing the galaxy-CMB lensing cross-power spectrum with \texttt{NaMaster}, it is possible to employ individual masks for each of the maps. For the DES field, we utilize the same mask used for computing the galaxy-power spectrum. For the $\kappa^{\rm CMB}$ field, we utilize the square of the mask described in Sec. \ref{sec:data}, while also setting the \texttt{masked\_on\_input} keyword to \texttt{True} in \texttt{NaMaster}. This approach is adopted because the reconstructed CMB lensing map inherently incorporates an apodized ACT mask, given that it is applied to the CMB maps during the quadratic estimator process. We validate the mask treatment in Sec. \ref{sec:pipeline_validation}.

 
When sky coverage is incomplete one needs to bin the pseudo-power spectrum and the coupling matrix. We compute the power spectra considering linearly spaced bins of width $\Delta \ell = 30$ up to $\ell= 3072$. However, in our analysis, we consider a conservative cut on values of $\ell$ smaller than  $\ell_{\rm min} = 50$ for the galaxy-power spectrum and $\ell_{\rm min} =100$ for the galaxy-CMB lensing power spectrum. We define these minimum multipoles based on the coverage of the sky and the precision of the lensing reconstruction \citep{darwish2021atacama}, respectively.\  

For modeling described in Sec. \ref{sec:th} to be accurate, we need to ensure that our analysis encompasses scales in the linear regime, where effects such as the non-linear galaxy bias and baryonic effects on the matter power spectrum are negligible. In the DES Y3 galaxy clustering analysis \citep{porredon2021dark_maglim} a scale cut corresponding to a comoving scale smaller than $R= 8 ~h^{-1}$\rm{Mpc} is imposed to exclude smaller scales on which the linear galaxy bias model breaks down. This cut corresponds to a $k_{\rm max}= 1/R = 0.125 ~h$\rm{Mpc}$^{-1}$. 
 
In our analysis, we define different scale cuts for $C_{\ell}^{gg}$ and $C_{\ell}^{\kappa g}$. For the galaxy power spectrum, we impose the scale cuts in each tomographic bin based on the ``physical scale cuts'', where $\ell_{\rm max} = k_{\rm max} \chi(\bar{z})$, with a conservative $k_{\rm max} = 0.08~h$\rm{Mpc}$^{-1}$, hence dropping physical scales below $12.5 ~h^{-1} \rm{Mpc}$. Since the galaxy-CMB lensing power spectrum is less affected by the non-linearities of the galaxy bias and has expected signal-to-noise smaller than the galaxy power spectrum, we impose a more flexible scale cut for $C_{\ell}^ {\kappa g}$, with $\ell_{\rm max}$ corresponding to the physical scales $k_{\rm max}$ fixed at $0.1~h\rm{Mpc}^{-1}$. These conditions led to the $\ell_{\rm max}$ values summarized in Table \ref{table:specifics}.

The galaxy power spectrum measurement has a shot noise contribution that must be subtracted. Assuming purely Poisson sampling, the ``mode-coupled'' shot noise can be estimated analytically \citep{garcia2021growth} as
\begin{equation}
     \tilde{N}_{\ell} = \frac{\langle m \rangle}{\bar{n}_2},
 \label{eq:shot-noise}
 \end{equation} 
where $m$ is the survey mask, and $\bar{n}_2$ is effective number density given by
\begin{equation}
    \bar{n}_2 = \frac{(\sum_{i\in \rm p}w_i)^2}{\Omega_{\rm pix}\sum_{\rm p}m_{\rm p}\sum_{i\in \rm p}w_{i}^2},
\end{equation}
where $\Omega_{\rm pix}$ is the pixel area in units of steradians. The ``mode-decoupled'' bandpowers are then computed by taking the inverse of the binned mode-coupling matrix 
\begin{equation}
N_{\ell} = \sum_{\ell'} (M^{-1})_{\ell\ell'} \tilde{N}_{\ell'}.  
\label{eq:shot-noise2}
\end{equation}

We test the validity of the shot-noise estimate in Sec. \ref{sec:shot_noise}. In the subsequent analysis, we subtract the shot-noise from the galaxy power spectrum $C_{\ell}^{gg}$, and then divide the bandpowers by the square of the pixel window function. The $C_{\ell}^{\kappa g}$ is divided by one power of the pixel window
function. Finally, when comparing data with theory, we apply the appropriate binned mode-coupling matrix and binning scheme to the theoretical curves.

\subsection{Covariance Matrix}
\label{sec:cov_matrix}

In order to quantify the correlation between bandpowers in our analysis, we consider the so-called \textit{disconnected} part of the covariance, that is, we assume that all fields are Gaussian distributed. We do not include the non-Gaussian covariance term in our analysis, since we consider only linear or weakly nonlinear scales, where the \textit{disconnected} terms of the covariance are dominant and the non-Gaussian corrections are expected to be small \citep{barreira2018accurate}. In addition, the super-sample covariance term, responsible for correlations on large scales, is largely subdominant for a survey like DES Y3, as shown in \citep{2021MNRAS.tmp.2208F}. This means that, for the purposes of this work, we can safely employ a Gaussian covariance.
   
In the absence of sky masks, the different harmonic modes are uncorrelated in the Gaussian covariance, so its computation for two fields, X and Y, reduces to the diagonal terms given by
\begin{equation}
  \mathrm{Cov}^{XY}_{\ell\ell'}= (\sigma^{XY}_{\ell})^2 \delta_{\ell\ell'},
\end{equation}
where the error $\sigma^{XY}_{\ell}$ can be computed using
\begin{equation}
    (\sigma^{XY}_{\ell})^2 = \frac{1}{(2\ell+1)} [(C_{\ell}^{XX})(C_{\ell}^{YY})+ (C_{\ell}^{XY})^2],
    \label{eq:knox_error}
\end{equation}
where the auto-power spectra terms above contain the contribution of the associated noise, $N_{\ell}$.
 
While the expression above is accurate for full-sky and some particular cases of partial sky \citep{knox1995determination}, a sky mask introduces a non-zero correlation between different modes in the Gaussian covariance and may also affect the amplitude of the main diagonal elements. To account for these, we use the method proposed in \cite{efstathiou2004myths, garcia2019disconnected}, implemented in \texttt{NaMaster}, where the covariance is computed as
\begin{equation}
    \rm{Cov}(C_{\ell}^{\rm XY,obs}C_{\ell'}^{\rm XY,obs}) = \frac{1}{(2\ell'+1)}[C_{(\ell}^{XX}C_{\ell')}^{YY}\mathrm{M}_{\ell\ell'}+C_{(\ell}^{XY}C_{\ell')}^{YX}\mathrm{M}_{\ell\ell'}].
\end{equation}
The coupling matrix $\mathrm{M}_{\ell\ell'}$ is computed based on the masks of fields $X$ and $Y$ (see \cite{garcia2019disconnected} for further details).

The \texttt{NaMaster} algorithm requires the use of the underlying power spectra,  $C_{\ell}^{gg}, C_{\ell}^{\kappa g}$ and $C_{\ell}^{\kappa\kappa}$, for example, from the theoretical prediction that depends on parameters that we do not know \textit{a priori}, such as the galaxy bias. In order to circumvent this issue and perform our analysis according to the blinding policy described in Sec. \ref{sec:blinding}, 
we compute the covariance using the power spectra computed with the fiducial cosmology described in Table \ref{tab:params}. After unblinding the data results, we update the analysis with the covariance computed with this best-fit cosmology, finding $\sim 0.3\sigma$ changes in the main result for $S_8$.

When computing the covariance, we consider the noise contribution of the galaxy field given by Eq. \ref{eq:shot-noise2}. For the CMB convergence field, we estimate the noise contribution by taking the residual between the power spectrum of the noisy ACT CMB lensing simulations \citep{darwish2021atacama} and the noiseless CMB lensing simulations used as input to lens the ACT CMB simulations. This approach takes into account the impact of the noise caused by the lensing reconstruction and survey geometry. As a sanity check, in section \ref{sec:cov_mocks} we use lognormal mock galaxy catalogs to check the robustness of the covariance matrix computation.%

\begin{table}
	\centering	
	\caption{	
		The parameters and their priors used in the analysis considering a flat $\Lambda$CDM universe. Square brackets denote a flat prior, while parentheses denote a Gaussian prior of the form $\mathcal{N}(\mu,\sigma)$. The Fiducial column shows the fiducial values we consider to construct the simulations in Sec \ref{sec:logmocks}. \\} 
	\vspace{-0.2cm}
	\begin{tabular}{ccc}
		\hline
		Parameter & Fiducial & Prior \\\hline
		\multicolumn{3}{c}{\textbf{Cosmology}}  \\
		$\Omega_{\rm m}$ &  0.309 &[0.1, 0.9] \\ 
		$A_\mathrm{s}10^{9}$ & 2.19 & [$0.5$, $5.0$]  \\ 
		$n_{\rm s}$ & 0.97 & [0.87, 1.07]  \\
		$\Omega_{b}$ & 0.049	&[0.03, 0.07]  \\ 
		$h_0$  & 0.69 &[0.55, 0.91]   \\
		\\\hline
		
		\multicolumn{3}{c}{\textbf{Linear galaxy bias  } } 	 \\
		$b_{i}$  & $1.5,1.8,1.8,1.9, 2.3, 2.3$ & [0.8,3.2]\\\hline
		
		\multicolumn{3}{c}{\textbf{Lens
				magnification } }   \\
		$s_{i} $ & 0.642, 0.63 , 0.776, 0.794, 0.756, 0.896 & Fixed\\ \hline
		
		\multicolumn{3}{c}{\textbf{Lens photo-z shift and stretch }}  \\
		$\Delta z_1 $ & 0.0 & ($
		-0.009, 0.007$)\\ $\Delta z_2 $ & 0.0 & ($
		-0.035, 0.011$)\\ $\Delta z_3 $ & 0.0 & ($
		-0.005, 0.006$)\\ $\Delta z_4 $ & 0.0 & ($
		-0.007, 0.006$)\\ $\Delta z_5$ & 0.0 & ($
		0.002, 0.007$)\\ $\Delta z_6$ & 0.0 & ($
		0.002, 0.008$)\\ $\sigma z_1$ & 1.0 & ($
		0.975, 0.062$)\\ $\sigma z_2$ & 1.0 & ($
		1.306, 0.093$)\\ $\sigma z_3$ & 1.0 & ($
		0.870, 0.054$)\\ $\sigma z_4$ & 1.0 &
		($0.918, 0.051$)\\ $\sigma z_5$ & 1.0 &
		($1.08, 0.067$)\\ $\sigma z_6$ & 1.0 &
		($0.845, 0.073$) \\\hline 
		
		\hline
	\end{tabular}
	\label{tab:params}

\end{table}

\subsection{Parameter inference}
\label{sec:methd_parameter}
In order to extract cosmological information from the measured $C_{\ell}^{gg}$ and the $C_{\ell}^{\kappa g}$, we evaluate the posterior of the parameters conditional on the data by assuming a Gaussian likelihood,  $\mathcal{L}$, of the form
 \begin{equation}
  \ln \mathcal{L}(\textbf{D}|\boldsymbol{\theta}) = -\frac{1}{2} [\textbf{D} - x(\boldsymbol{\theta})]^{\mathrm{T}}\mathrm{Cov}^{-1}[\textbf{D} - x(\boldsymbol{\theta})],
  \label{eq:likelihood}
 \end{equation}
 where $\textbf{D}$ is the measured data vector, $x(\boldsymbol{\theta})$ is the theoretical prediction at parameter values $(\boldsymbol{\theta})$. The quantity $\mathrm{Cov}^{-1}$ denotes the inverse of the covariance matrix, described in Sec \ref{sec:cov_matrix}, and which we keep constant during parameter estimation. The posterior distribution is then proportional to the product of the likelihood and the priors
 \begin{equation}
     p(\boldsymbol{\theta}|\textbf{D}) \propto \mathcal{L}(\textbf{D}|\boldsymbol{\theta})\pi(\boldsymbol{\theta}),
 \end{equation}
where $\pi(\boldsymbol{\theta})$ are the priors on the parameters of our model.

We consider a spatially flat $\Lambda$\rm{CDM} cosmology, characterized by the total matter density $\Omega_{m}$, the baryonic density $\Omega_{b}$, the dimensionless Hubble parameter $h$, the amplitude of primordial scalar density perturbations $A_{s}$, and the spectral index $n_{s}$ of the power spectrum.  We do not expect massive neutrinos to impact our analysis, given our scale cuts, the expected significance of the cross-correlation, and that DES Y3 clustering analysis poorly constrains the neutrino mass \citep{descollaboration2021dark}. Therefore, we consider only massless neutrinos to speed up the computation of the matter power spectrum. The prior ranges for the parameters are listed in Table \ref{tab:params}, and are motivated by physical constraints or by the DES Y3 analysis \citep{porredon2021dark_maglim}.

We sample the likelihood using the Monte Carlo Markov Chain (MCMC) sampler, implemented in the publicly available code \texttt{Cobaya}\footnote{\url{https://cobaya.readthedocs.io/en/latest/index.html}}\citep{torrado2019cobaya,lewis2013efficient}. We determine the chain convergence using a generalized version of the $R-1$ Gelman-Rubin statistic \citep{lewis2013efficient,gelman1992inference}, which we establish that the chains converge once  $R -1 < 0.05$. We remove the first 30$\%$ of the chains from all analyses as burn-in. All the visualization of our results is done with GetDist\footnote{\url{https://getdist.readthedocs.io/en/latest/}} \citep{lewis2019getdist}.

\subsection{Blinding}
\label{sec:blinding}
The results were blinded throughout the analysis until we tested the pipeline and passed all the tests for systematic effects. The tests we performed before unblinding the cosmological results are summarized below:
\begin{itemize}
    \item First, we tested all the stages of the analysis pipeline using simulations, described in Sec. \ref{sec:test_logmocks}. We validated that we could recover the input power spectra, cosmological, and nuisance parameters of the simulations.  
    \item  Next, we used the real data to test for possible systematic contamination in our measurements. These include: 1) check the robustness of the weights applied to correct correlations in the number density map with survey properties; 2) understand and test for spurious contributions to our $C_{\ell}^{\kappa g}$ signal due possible correlated foregrounds; 3) validate the shot-noise subtraction in the $C_{\ell}^{gg}.$
    \item Once the above issues were well understood, we ran the parameter inference using real data under different choices in the pipeline. In this step, we looked at the blinded contours only, removing the axes when generating the figures of the parameter constraints. The real data points were not directly compared with theoretical predictions until they were unblinded. 
\end{itemize}

We describe the changes done post-unblinding in Sec. \ref{sec:unblind}.

\section{SIMULATIONS AND PIPELINE VALIDATION}
\label{sec:test_logmocks}

In this section, we apply our methodology to mock simulations to assess the validity of the power spectrum computation, covariance estimation, and the parameter constraints over the range of angular scales considered.

\subsection{Lognormal Mocks}
\label{sec:logmocks}
An important step before unblinding is to validate the pipeline using synthetic data with the noise levels, survey geometry, and galaxy density as in the real data. While the variance of the galaxy-CMB lensing convergence power spectrum can be captured to sufficient accuracy by Gaussian random fields, this is not realistic for the galaxy power spectrum due to the non-linear evolution of the density field \citep{coles1993testing}. On the scales we consider in our analysis, the one and the two-point function of the galaxy field can be better modeled by a field drawn from a lognormal distribution \citep{coles1991lognormal,wang2011perturbation,neyrinck2009rejuvenating,clerkin2017testing}.\

We follow the methodology described in Sec. 3.1 of \cite{carron2012inadequacy} to produce a set of 1000 correlated Gaussian and lognormal simulations mimicking the ACT CMB lensing field and the DES galaxy overdensity, respectively \footnote{The code used to produce the correlated mocks is publicly available at \url{https://github.com/huffenberger-cosmology/lognormal_mocks}}. For this, we consider the mean galaxy number density per steradian of the real data and the theoretical power spectrum computed based on the cosmological parameters, linear galaxy bias, lens magnification, and photometric redshift uncertainties listed in the ``Fiducial'' column of Table \ref{tab:params}.

We draw random galaxies following a Poissonian distribution to add the shot-noise within the survey's footprint. Then, we convert the galaxy number counts maps into galaxy overdensity maps $\delta_{g}^{\rm mock}$. Finally, we add Gaussian realizations of the CMB lensing noise \citep{darwish2021atacama} to the CMB lensing simulations. We account for the masks and the same pixelization scheme described in Sec \ref{sec:masks_maps} in the mock production.

\subsection{Angular power spectrum validation}
\label{sec:pipeline_validation}
 
\begin{figure}[ht]
    \centering
    \includegraphics[scale=0.8]{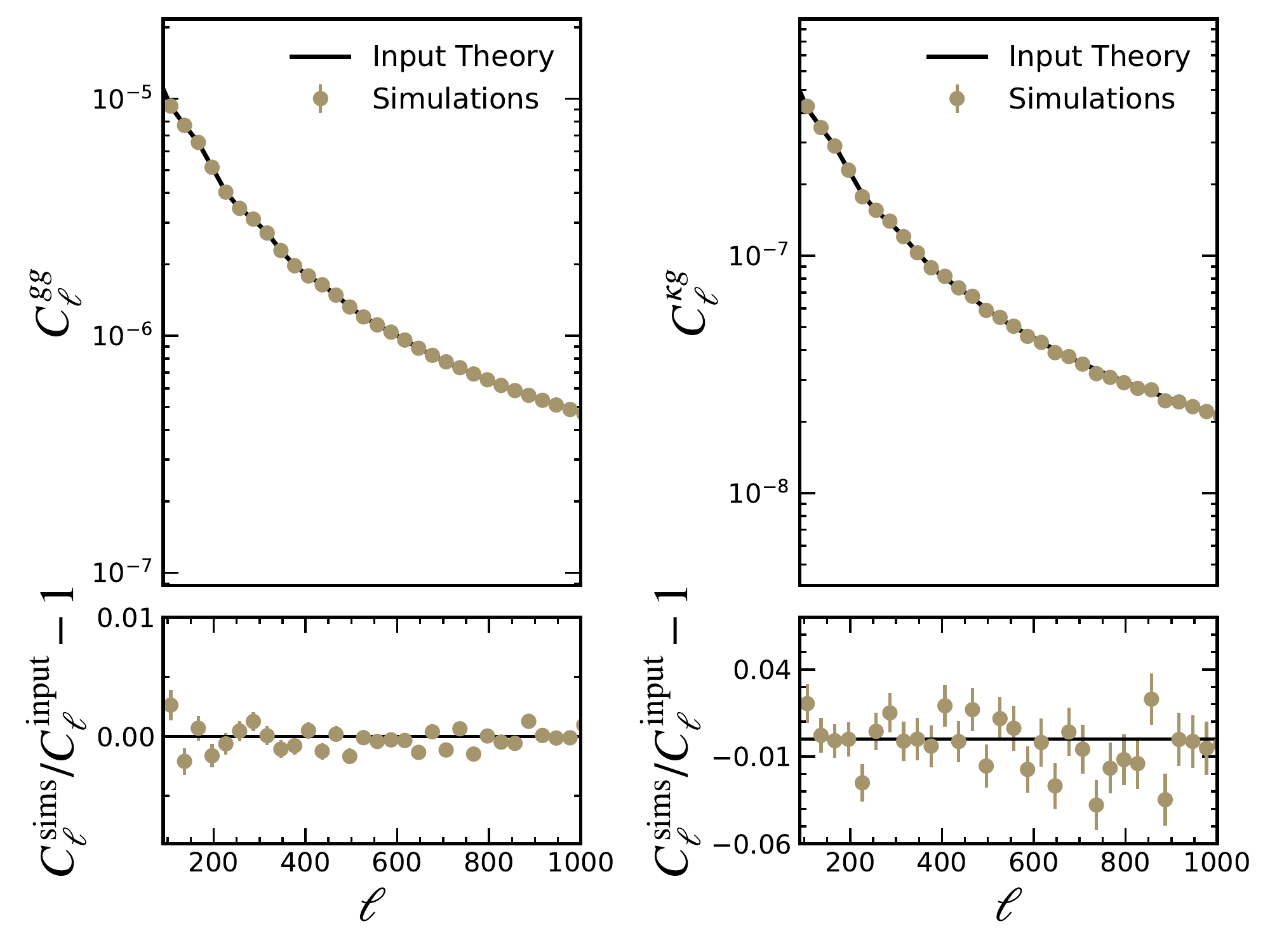}
    \caption{Comparison between the average power spectrum computed from simulations with the input power spectrum for the galaxy (left panel) and galaxy-CMB lensing (right panel) power spectrum. The error bars are divided by $\sqrt{N_{\rm{sims}}}$, where $N_{\rm{sims}}=1000$. We show the results for the second redshift bin, but are similar for all redshift bins. }
    \label{fig:cls_mocks}
\end{figure}

In order to validate the mask treatment in the power spectrum computation, we apply the methodology described in Sec. \ref{sec:methods} on the 1000 lensing convergence and galaxy overdensity mock simulations with known cosmology. In Fig. \ref{fig:cls_mocks} we show the average of the $C_{\ell}^{gg}$ (left panel) and $C_{\ell}^{\kappa g}$ (right panel) measured from these mocks, compared with the input theoretical power spectrum. We verify that both measurements, the auto and cross-power spectrum, are in good agreement with the input signal with sub-percent accuracy over the range of scales of interest. The level of agreement is much better than the measurement uncertainty for any of the bins. For clarity, the figure shows the result of the second redshift bin of the \maglim sample simulations, but these results are consistent for all redshift bins. 

\begin{figure}[ht]
    \centering
    \includegraphics[scale=0.7]{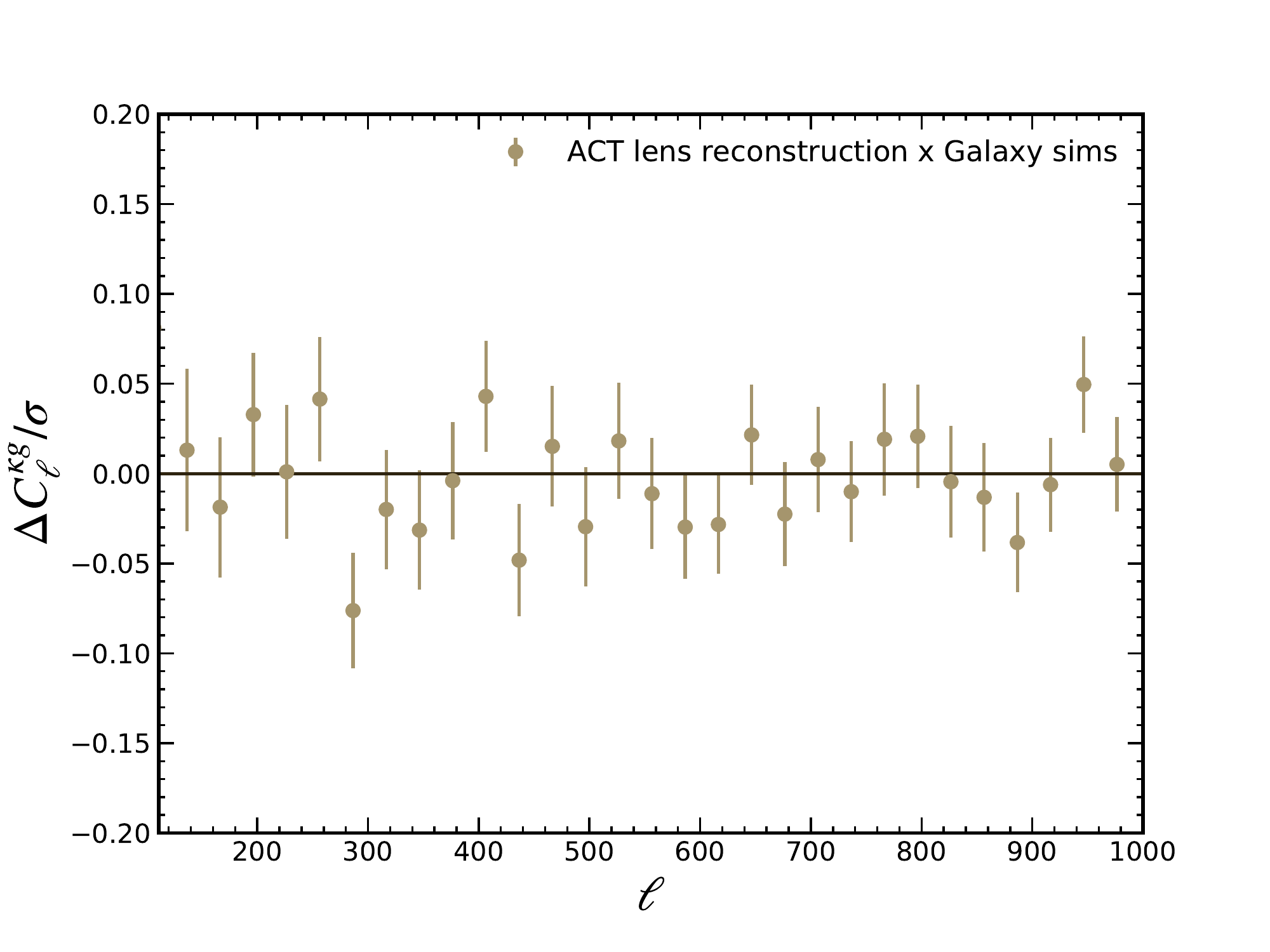}
    \caption{Test of any residual bias in the galaxy-CMB lensing power spectrum from lensing reconstruction and mask treatment. The points show the difference between the cross-power spectrum of the constrained realizations of DES galaxies and the ACT lensing simulation compared to the input, in units of the error bar of the data.}
    \label{fig:constrained_realization_gals}
\end{figure}

In addition to the mask treatment in the $C_{\ell}^{\kappa g}$ computation, we also want to test additional effects that may impact the ACT convergence reconstruction map and thus potentially bias our measurements. Such sources of bias could include, for example, the \texttt{HEALPix} projection of the convergence map, the mask treatment, the normalization correction in the lensing reconstruction, or any non-trivial mask effect during the lensing reconstruction. To this end, we generated 500 constrained realizations of the galaxy overdensity, which are correlated with the noiseless lensing potential maps used as the input to lens the ACT CMB simulations \citep{darwish2021atacama}. Then, we computed the cross-power spectrum of these galaxy simulations with the respective noisy suite of ACT lensing maps. These noisy lensing simulations accurately reflect the signal and noise properties of the ACT data. The $C_{\ell}^{\kappa g}$ obtained from this simulation suite is in agreement with the input power spectrum to within $< 0.1\sigma$ (where $\sigma$ here is the error of the mean), as shown in Fig. \ref{fig:constrained_realization_gals}.

\subsection{Covariance validation}
 \label{sec:cov_mocks}
In our analysis, the covariance matrix is estimated with the \texttt{NaMaster} algorithm, assuming a Gaussian (disconnected) approximation. When doing the parameter inference using multiple tomographic bins, we do not consider the cross-spectra between the tomographic bins. Nevertheless, we include all the cross-terms in our covariance, i.e., the cross-covariance of $C_{\ell}^{g_{i}g_{j}}$ and $C_{\ell}^{\kappa g_{i}}$. However, the resulting matrix is dominated by its diagonal components.

\begin{figure}[h!]
\centering
\includegraphics[scale=0.6]{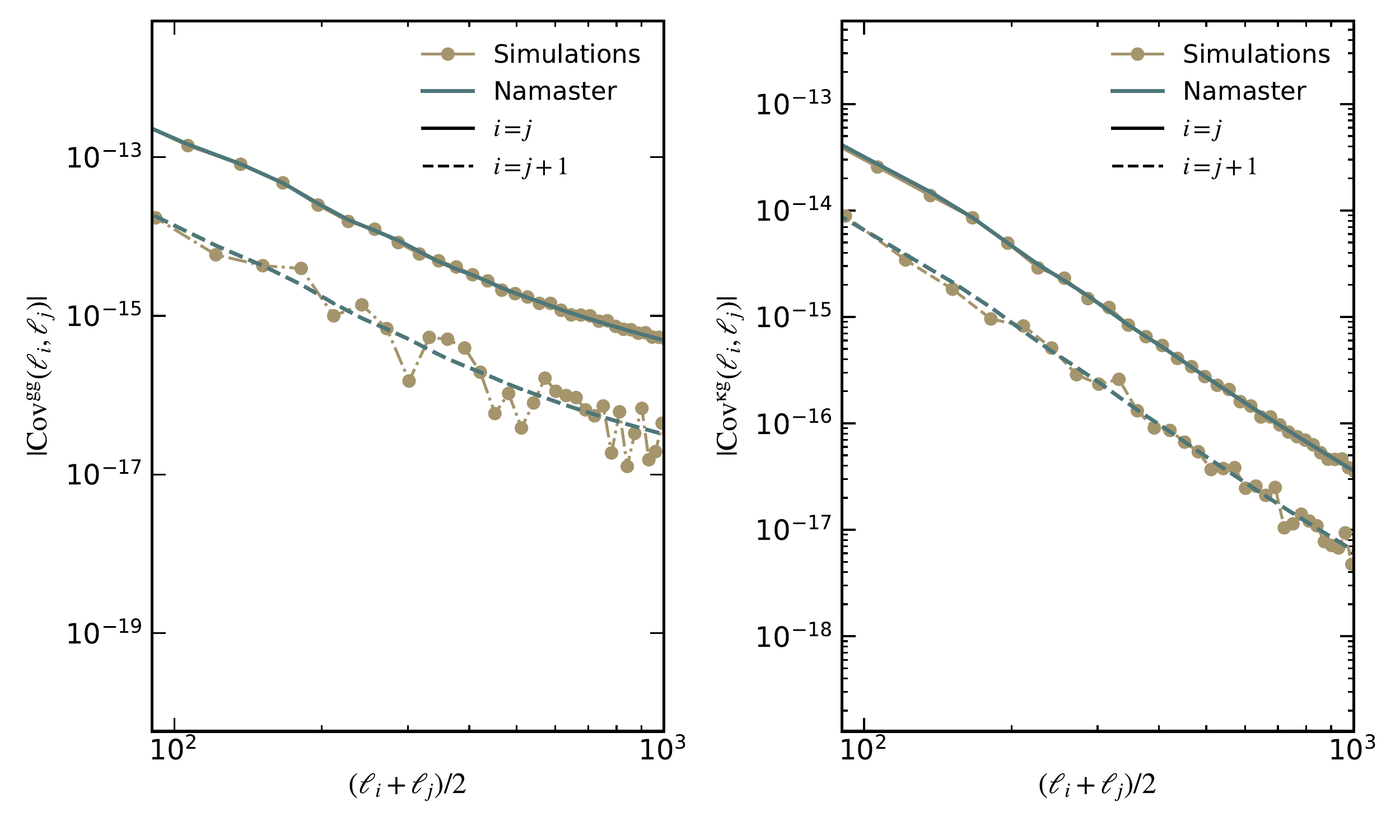} %
\caption{Comparison of the diagonal terms of the covariance matrix computed directly from the simulations and \texttt{NaMaster} covariances (see Section \ref{sec:cov_mocks}). The solid and dashed lines show the zeroth-order ($i = j$) and first-order ($i = j+1$) diagonals, respectively. For clarity, we show the covariance elements of $C_{\ell}^{gg}$ (left panel) and $C_{\ell}^{\kappa g}$ (right panel) of the fifth redshift bin, but we find similar agreement for the other bins.}%
\label{fig:sigma_cov_mocks}%
\end{figure}

We checked our covariance matrix against one produced by the lognormal simulations. Fig. \ref{fig:sigma_cov_mocks} shows the diagonals of the covariance matrix estimated using simulations and \texttt{NaMaster}. For both the galaxy auto-spectrum (upper panel) and galaxy-CMB lensing power spectrum (lower panel),  we find that the \texttt{NaMaster} covariance is in very good agreement with the covariance computed directly using the simulations. For clarity, we show the result of the fifth redshift bin, but we find analogous agreement for the other bins. Thus, we conclude that the \texttt{NaMaster} covariance accurately recovers the power spectrum uncertainties given the impact of the survey geometry.

\subsection{Parameter estimation on mocks}
We seek to constrain cosmological parameters using observations. In order to validate the parameter inference, we test the ability in recovering the input cosmological and nuisance parameters using simulations. We verify the results using the power spectra measured from two independent realizations as the data vector and also using the average of the set of simulations.  
 
\begin{figure}[ht]
    \begin{center}
    \includegraphics[scale=0.9]{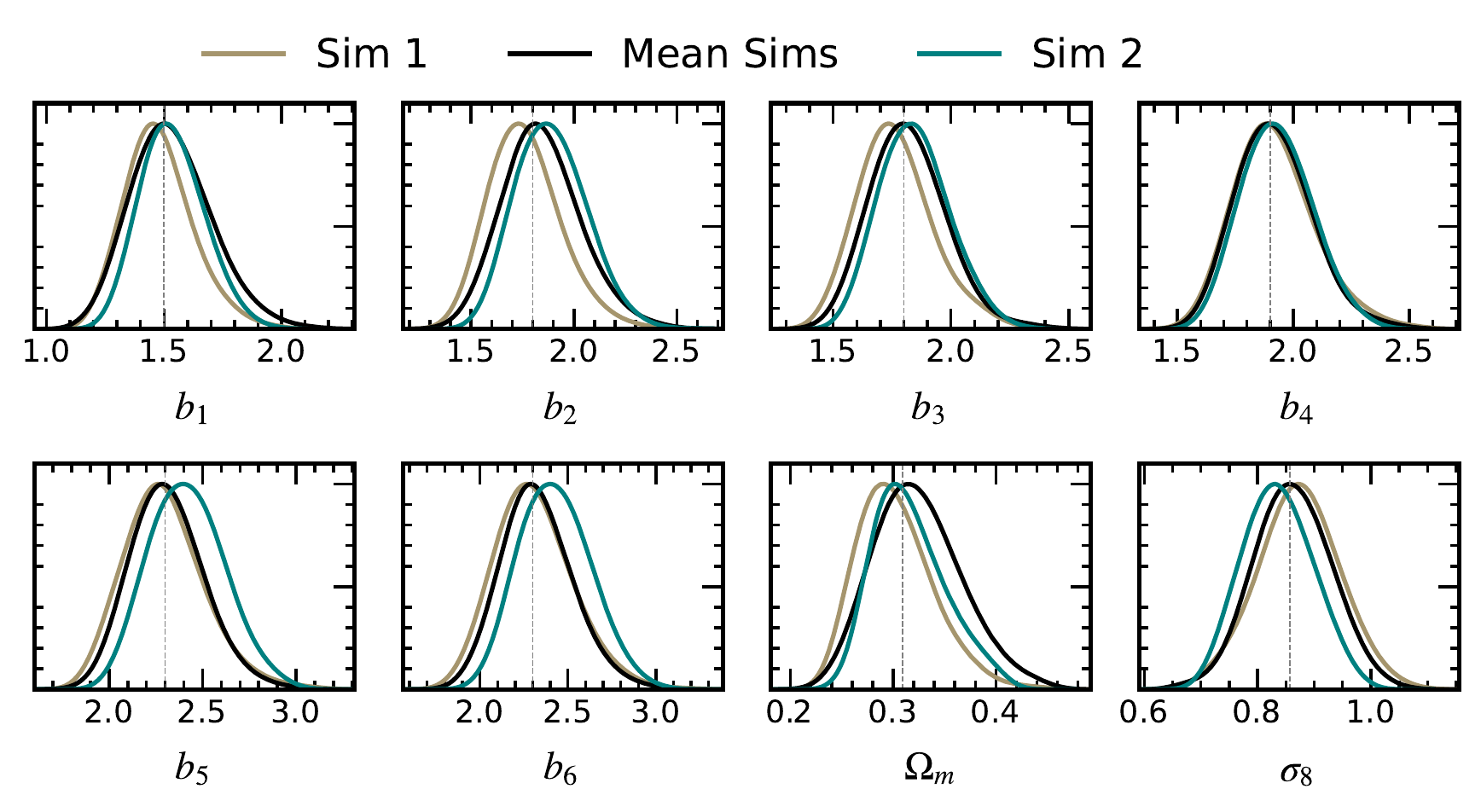}
    \caption{Synthetic parameter constraints obtained using simulations to test the pipeline.
     The solid curves show the marginal posterior distributions for 6 linear galaxy bias, $\sigma_{8}$ and $\Omega_{m}$ given a set of $C_{\ell}^{gg}$ and $C_{\ell}^{\kappa g}$ extracted from two independent realizations (colored lines) and from the average of the set of simulations (black curves). The dashed vertical lines depict the true values used in the simulations. The constraints from the data appear in Fig. \ref{fig:s8_baseline}.}
    \label{fig:logmocks_parameters}
    \end{center}
\end{figure}

We performed a $C_{\ell}^{gg}$ and $C_{\ell}^{\kappa g}$ joint fit for the 6 photo-z bins with a total of 23 free parameters, including the galaxy bias, photo-z uncertainties, and cosmological parameters. We impose the priors shown in Table \ref{tab:params} for the linear galaxy bias and cosmological parameters. As our simulations do not include the photo-z uncertainties, we consider a Gaussian prior with mean-centered on zero and one respectively for the photo-z shift and stretch parameters, and width equal to the values considered in the real data shown in Table \ref{tab:params}. We apply the same scale cuts as in the analysis of the real data.
In Figure \ref{fig:logmocks_parameters} we show the marginal posterior distributions for the linear galaxy bias of the six tomographic bins, $\Omega_{m}$ and $\sigma_{8}$ obtained for a given set of $C_{\ell}^{gg}$ and $C_{\ell}^{\kappa g}$ extracted from two independent realizations (yellow and blue curves) and from the average of the set of simulations (black curves). In all cases, the results are compatible with their input values, thus validating our pipeline. The small deviation from the input values seen in some of the parameters constrained from individual simulations is expected due to statistical fluctuations.

\section{ROBUSTNESS OF THE RESULTS}
\label{sec:sys}

The Y3 \maglim sample and the ACT DR4 CMB lensing map have undergone thorough tests for various astrophysical and systematic effects in previous studies \cite{rodriguez2021dark,aiola2020atacama,choi2020atacama,darwish2021atacama}. Here, we perform additional tests to further assess the robustness of our measured bandpowers.

\subsection{Weights vs deprojection}
Sky contaminants such as non-cosmological sources, artifacts in the images, and other systematic effects coming from the observing conditions may change the selection function across the observed footprint or with redshift. Consequently, the observed number density of galaxies may vary according to the conditions of the survey. A detailed analysis of the impact of these survey properties or observational effects in the \maglim number density was explored in \cite{rodriguez2021dark}, which derived weights to account for correlations in the number density with survey properties. As discussed in \ref{sec:des_data}, we apply these weights in our analysis when producing the total pixelized number counts map. A full characterization of the effect of observational systematics on the \maglim number counts is presented in \cite{rodriguez2021dark}. Here, we perform additional tests to ensure that the systematics and the survey properties do not significantly impact our measurements.

In order to validate that the weighting applied to the number counts map is robust to correct for the effect of systematics on the galaxy-power spectrum and on the galaxy-CMB lensing power spectrum, we apply a technique called “template deprojection” \citep{elsner2016unbiased, alonso2019unified}. For that, it is assumed that the observed galaxy overdensity, $\delta^{\rm obs}$, is modeled linearly with the contaminants so that  
\begin{equation}
    \delta^{\rm obs} = \delta^{\rm true}+\alpha t,
    \label{eq:deprojec_eq}
\end{equation}
where the "true" overdensity, $\delta^{\rm true}$, receives contributions from the template maps of the fluctuation of known contaminants, $t$, scaled by an unknown linear amplitude $\alpha$. The modes that are common to the set of systematic templates and the observed maps are removed by building a likelihood for $\delta^{\rm obs}$, considering a model for $\delta^{\rm true}$ and marginalizing over $\alpha$. This can be achieved by projecting $\delta^{\rm obs}$ onto the subspace orthogonal to $t$ and analytically accounting for the associated loss of modes when estimating the angular power spectrum. We refer the readers to the listed references for more details on the template deprojection method.

We use a set of 107 maps corresponding to DES survey properties (SP) in different photometric bands $\{ g,r,i,z\}$, stellar density, and the interstellar extinction map as the contaminant templates. This set corresponds to the systematics considered in the DES Y3 Key Project. More detailed information on the construction of these maps may be found in \cite{sevilla2021dark} and in \cite{rodriguez2021dark}.

To evaluate the impact of these contaminants, we compare the resulting power spectra after the template deprojection from galaxy overdensity maps constructed without the weights applied to the number counts. The SP deprojection and weighting are different techniques that supposedly correct for the same effects, so we expect the deprojected power spectra of unweighted galaxy maps to be consistent with the power spectra of the weighted maps.  

\begin{figure}[ht]
    \centering
    \includegraphics[scale=0.4]{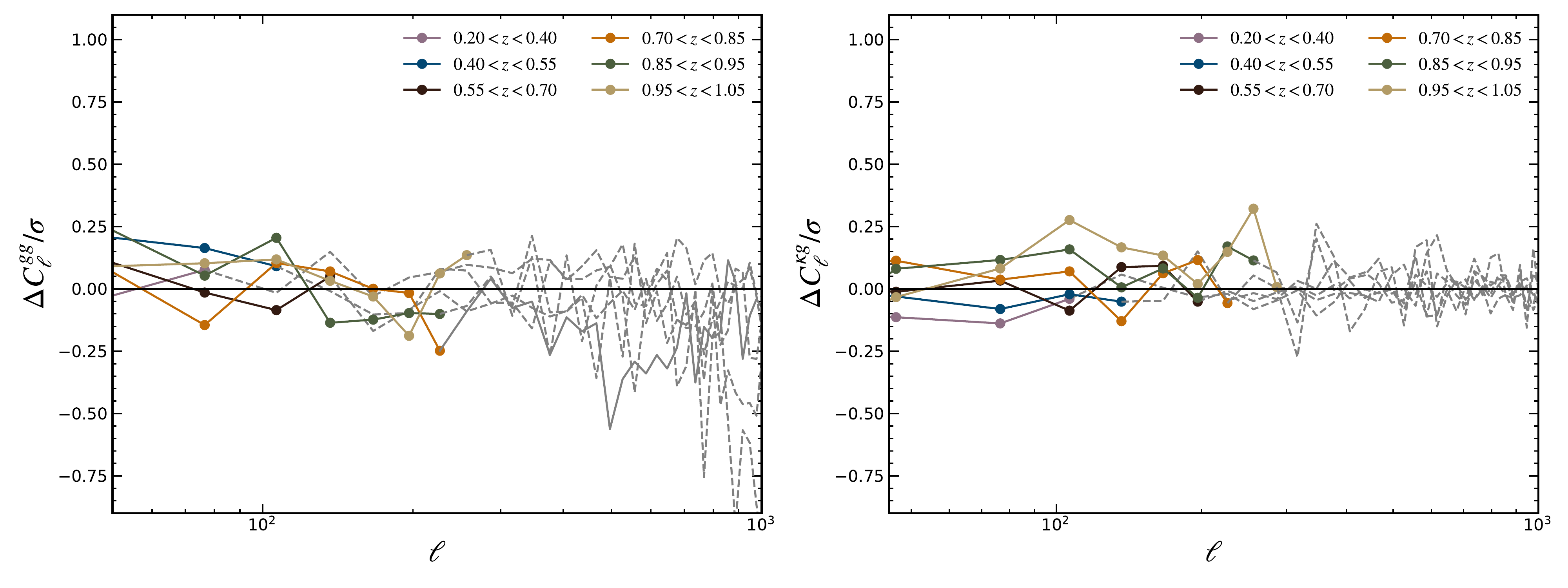}
    \caption{Comparison of the power spectrum measurements of weighted maps (our baseline measurements) and after deprojecting the contaminants of unweighted maps, in units of statistical error. For both measurements, galaxy auto-power spectrum (left panel) and CMB lensing-galaxy power spectrum (right panel), the residual is less than $\sim 0.3\sigma$ within the scales of interest (colored lines), suggesting no significant contamination. The grey dashed lines denote the multipole range not used in our analysis. }
    \label{fig:deprojection}
\end{figure}

Fig. \ref{fig:deprojection} shows the difference between the power spectrum measurements of weighted maps (our baseline measurements) and after deprojecting the contaminants of unweighted maps, in units of the statistical error. In both measurements, $C_{\ell}^{gg}$ (left panel) and $C_{\ell}^{\kappa g}$ (right panel), the deprojected $C_{\ell}$ of unweighted maps are only sub-percent different from the baseline measurements, by less than $\sim 0.3\sigma$ within the scales of interest. This consistency provides additional confidence that the \textit{SP} are already accounted for by the data cuts and weighting since their effects are subdominant with respect to other sources of uncertainty.

\subsection{tSZ and CIB effects}
One of the major contaminants to the CMB lensing cross-correlations with tracers of large-scale structure is the tSZ effect. For this reason, in our baseline analysis, we use a CMB lensing map where tSZ has been explicitly removed (``tSZ-free''). To evaluate the impact of the tSZ effect on the CMB lensing map on our analysis, we perform an additional test by measuring the cross-correlation between the \maglim galaxies and a CMB lensing map without tSZ removal (``with-tSZ''). The aim of this test is to characterize the change in the $C_{\ell}^{\kappa g}$, as well as any potential scale dependence caused in the presence of the tSZ.
 
To assess the tSZ effect, we reconstruct the with-tSZ map based on ACT-only data, instead of the combination of ACT and Planck's data used in the tSZ-free map \citep{darwish2021atacama}. These two maps contain slightly different noise levels. We estimate the noise level of the with-tSZ map by taking the mean difference of the auto-power spectrum of 500 ACT simulations and their input power spectrum. These simulations are built considering the same setup and reconstruction method used for the data. We use the associated noise of the field and the method described in Sec. \ref{sec:cov_matrix} to compute the covariance of the $C_{\ell}^{\kappa^{\rm with-tSZ} g}$.

\begin{figure}[ht]
    \begin{center}
    \includegraphics[scale=0.6]{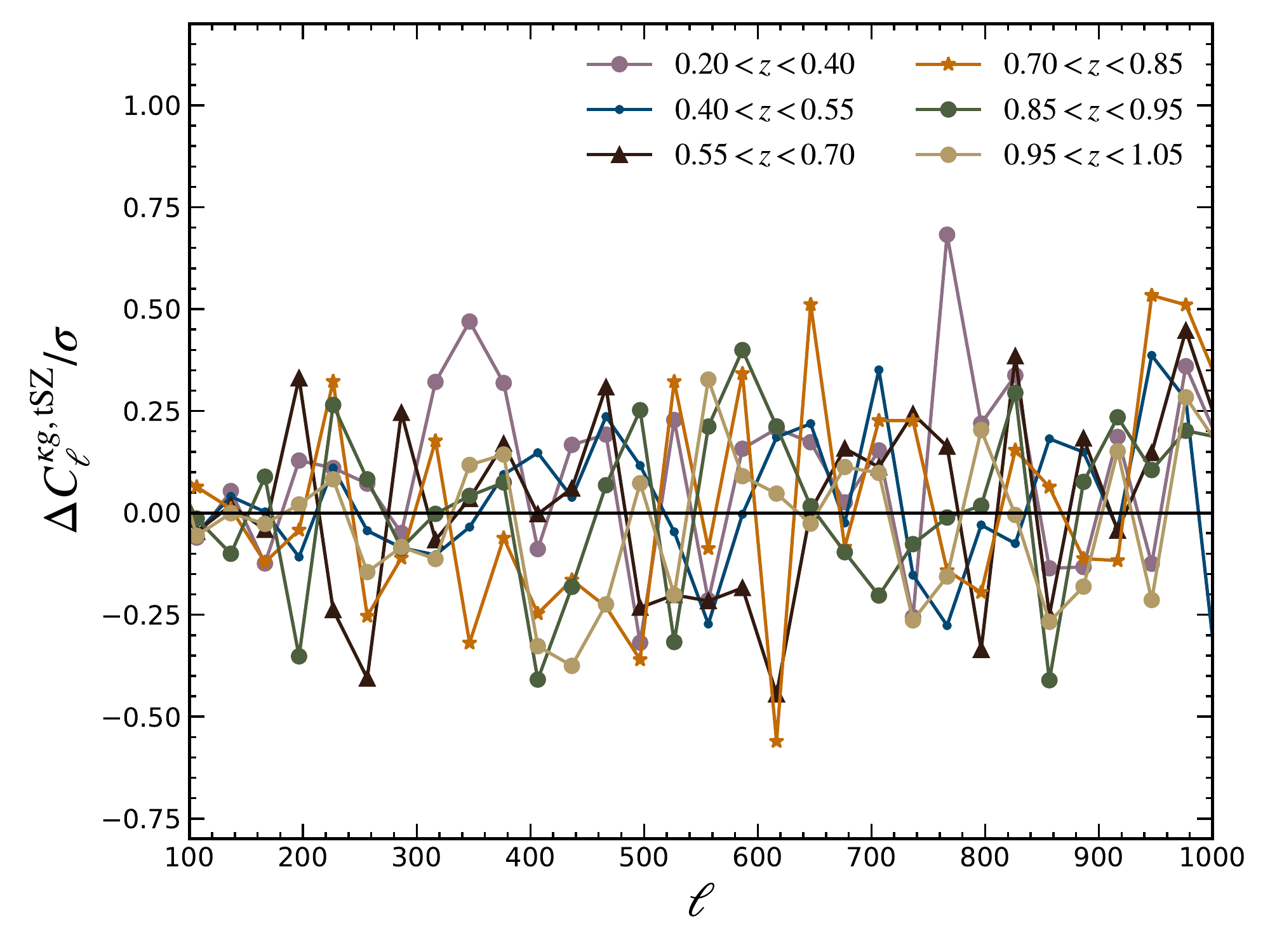}
    \caption{Difference between the $C_{\ell}^{\kappa g}$ extracted from the convergence map with tSZ and from the tSZ-free map, in units of the total error of the measurement. Although we use the tSZ-free convergence map in our baseline analysis, the power spectrum does not change significantly if we do not deproject the effect, by less than $\sim 0.3\sigma$ in the scales of interest.}
    \label{fig:cl_tsz_test}
    \end{center}
\end{figure}

Figure \ref{fig:cl_tsz_test} shows the difference between the $C_{\ell}^{\kappa g}$ extracted from the convergence map with tSZ and from the tSZ-free map, in units of the total error of the measurement. Although we use the tSZ-free convergence map in our baseline analysis, the power spectrum does not change significantly if we do not deproject the effect, by less than $\sim 0.3\sigma$ in the scales of interest and with no clear scale dependency. Therefore, we can conclude that even if we had used a CMB lensing map without tSZ removal, it would not significantly impact our results.
 
Another expected bias to the galaxy-CMB lensing cross-correlations is the CIB originated dominantly from the emission from unresolved dusty star-forming galaxies. At the same time that the CIB correlates with the galaxies, its signal may leak into convergence in the lensing reconstruction \citep{mishra2019bias}. However, we do not expect a strong correlation between CIB with the \maglim galaxy sample since the CIB emission is mostly sensitive to redshifts $z\sim 2$, higher than those probed by DES galaxies. The impact of the CIB in the galaxy-CMB lensing cross-correlation considering DES-like galaxies was explored in \cite{baxter2019dark}, finding that the CIB contamination is sub-dominant compared to the tSZ contamination (see. Fig 3 of \cite{baxter2019dark}). Since our results are stable in the case of a CMB with tSZ (as shown in Fig. \ref{fig:cl_tsz_test} ), we conclude that any possible bias introduced by the CIB is insignificant compared to our uncertainties.

\subsection{Null-tests}
\label{sec:null}

\begin{figure}[ht]
    \begin{center}
    \includegraphics[scale=0.6]{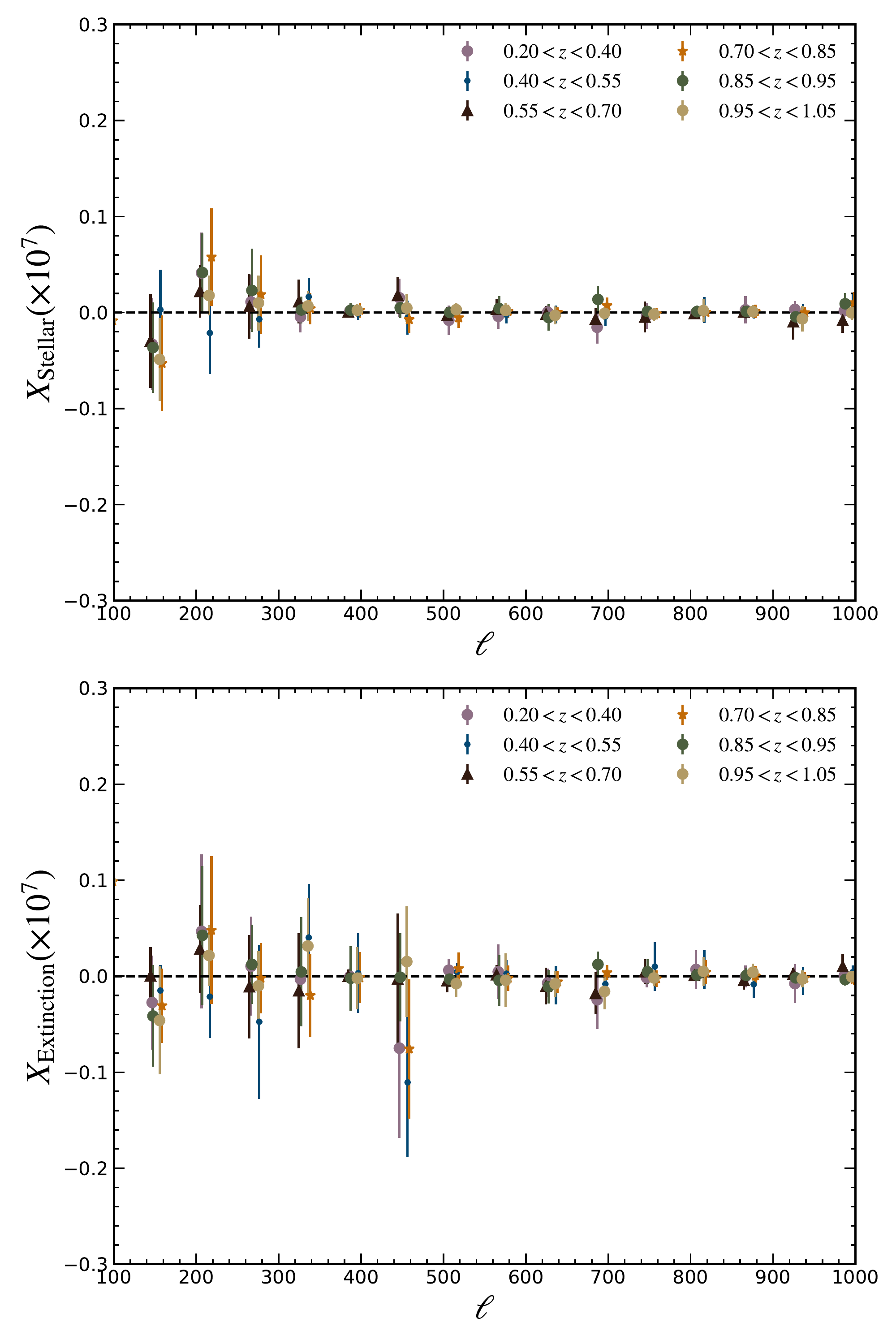}
    \caption{Contributions due to stellar contamination (upper panel) and extinction (lower panel) to the galaxy-CMB lensing cross-power spectrum, assessed by Eq. \ref{eq:Xsys}. The error bars are computed from the covariance obtained from a jackknife samples. In all cases we find no significant signal, and therefore, the null-test is consistent with zero. For clarity, the points are slightly shifted along the x-axis.  }
    \label{fig:clksystematics}
    \end{center}
\end{figure}
Cross-correlation between independent probes is generally less sensitive to known (and unknown) additive systematics compared to auto-correlation. However, there are systematics that can contaminate both, CMB lensing and galaxy overdensity and therefore, impact our $C_{\ell}^{\kappa g}$. For instance, the extinction of distant sources by dust in our galaxy can modulate the number counts of galaxies and add signals in CMB temperature and polarization data. The stellar density is a potential source of systematic error, as it can correlate with foregrounds that might bias lensing reconstruction and modulate galaxy number density, potentially leading to a dilution of both auto and cross-correlations \citep{krolewski2020unwise}. In order to check the validity of our measurements against the possibility of spurious signals from contamination we compute 
\begin{equation}
X_{S} = \frac{C_{\ell}^{\kappa S}C_{\ell}^{\delta_{g}S}}{C_{\ell}^{SS}}, 
\label{eq:Xsys}
\end{equation}
where $S$ denotes the systematic map, either the stellar contamination or dust extinction.  The $X_{S}$ quantity accounts for the cross-power spectrum between the systematic map and both, $\kappa$ and $\delta_{g}$, normalized by the auto-power spectrum of $S$. If our measurements are not biased by these contaminants, we would expect this signal to be consistent with zero, compared to the statistical uncertainty of the measurements. The star map is created with bright DES point sources, hereafter \textit{stellar\_dens} \citep{sevilla2021dark}, and the interstellar extinction map is made following \cite{schlegel1998maps}.

We find no evidence for contamination and Fig. \ref{fig:clksystematics} shows $X_{S}$ of the stellar density (upper panel) and dust extinction (lower panel) for all redshift bins. For clarity, the points are slightly shifted along the x-axis. The error bars are computed using the ``delete one jackknife'' (JK) covariance \citep{shao1986discussion}, computed using $N_{\rm{jk}} = 37$ equal-area patches. Since the number of removable JK-patches is limited by the fraction of the sky, we reduced the number of $\ell$-bins when computing $X_{S}$ to obtain a more stable covariance. We have measured the $X_{S}$ with linearly spaced bins of a width of $\Delta\ell= 60$, instead of the $\Delta\ell=30$ used for the baseline analysis. We use the jackknife covariance to estimate the $\chi^2$ with respect to the null model ($X_{S}$ = 0), and the corresponding probability-to-exceed (PTE) for the same scale cuts summarized in Table \ref{table:specifics}. The results for each case are summarized in Table \ref{tab:null}. In all redshift bins, we find no significant signal, and therefore, the results are in agreement with the hypothesis of no systematic contribution in  $C_{\ell}^{\kappa g}$.

\begin{table}[ht]
\centering
\begin{tabular}{cccc}
\hline
\textbf{Correlation} & \textbf{Photo-z bin}                 & \textbf{$\chi^2$/$d.o.f$} & \textbf{PTE ($\%$)} \\ \hline
Stellar Density      & $0.20 < z< 0.40$                     & 0.46/1                    & 49.3               \\
                     & $0.40 < z< 0.55$                     & 0.005/1                   & 94.2               \\
                     & $0.55 < z< 0.70$                     & 0.36/2                    & 83.3               \\
                     & $0.70 < z< 0.85$                     & 1.31/2                    & 51.9               \\
                     & $0.85 < z< 0.95$                     & 1.22/3                   & 74.7             \\
\multicolumn{1}{l}{} & \multicolumn{1}{l}{$0.95 < z< 1.05$} & 1.99/3                    & 57.3               \\ \hline
\textbf{Correlation} & \textbf{Photo-z bin}                 & \textbf{$\chi^2/d.o.f$}   & \textbf{PTE ($\%$)} \\ \hline
Extinction           & $0.20 < z< 0.40$                     & 0.31/1                    & 57.4               \\
                     & $0.40 < z< 0.55$                     & 0.31/1                    & 57.4               \\
                     & $0.55 < z< 0.70$                     & 0.05/2                    & 97.3               \\
                     & $0.70 < z< 0.85$                     & 0.63/2                    & 72.6               \\
                     & $0.85 < z< 0.95$                     & 0.68/3                    & 87.7               \\
\multicolumn{1}{l}{} & $0.95 < z< 1.05$                     & 0.84/3                    & 83.9             
\end{tabular}
\caption{Summary of $\chi^2$ per degree of freedom and the PTE for the null-test defined in Eq. \ref{eq:Xsys}. The top half of the table shows the results for the stellar density, while the lower half shows the corresponding values for the extinction map.}
\label{tab:null}
\end{table}

\subsection{Shot-noise subtraction}
\label{sec:shot_noise}
As discussed in Sec. \ref{sec:method_PS}, we subtract the shot-noise contribution to the auto-correlation power spectra using an analytical estimate given by Eq. \ref{eq:shot-noise}. However, the shot-noise can deviate from this simple relation due to several effects such as super-Poissonian variance, variations in completeness, mask, and observational systematics \citep{baldauf2013halo}.\

\begin{figure}[ht]
 \begin{center}
    \includegraphics[scale=0.6]{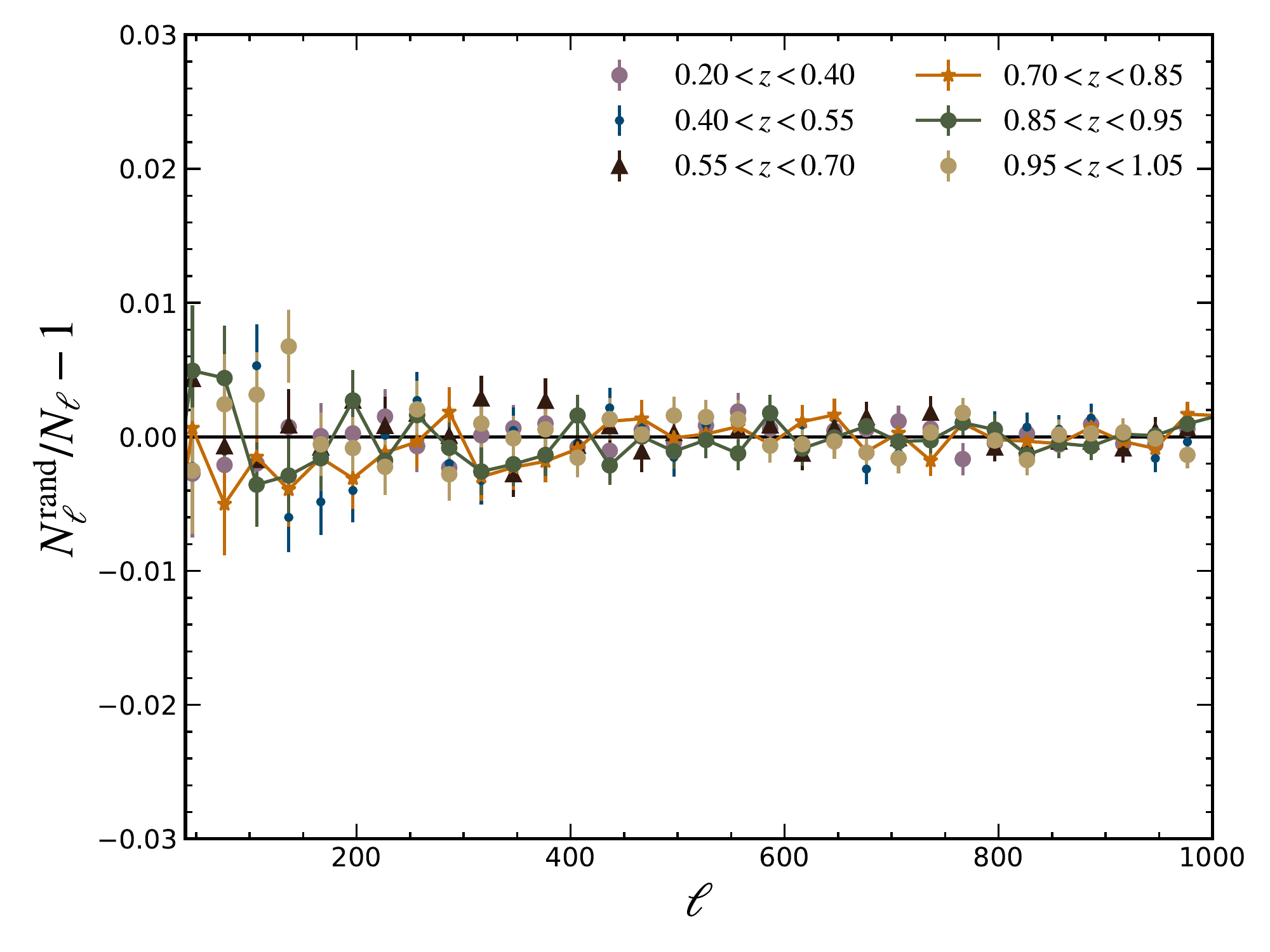}
    \caption{Comparison of the shot noise estimated analytically from Eq. \ref{eq:shot-noise} and \ref{eq:shot-noise2} and from the mean of random overdensity maps with the same number of objects as the real data. The analytic estimate agrees within $\sim 1\%$ with the random estimate for all redshift bins in the scales of interest.}
    \label{fig:shotnoise}
\end{center}
\end{figure}

We verify the validity of the analytical shot-noise by taking the power spectrum of random overdensity fields. To generate these fields, we randomized positions according to the completeness map, considering the same number of objects as the real galaxy sample. In essence, these randomized fields represent an independent Poisson process sampling the same smooth overdensity field. Fig. \ref{fig:shotnoise} shows the comparison between the shot-noise from the analytical prediction with the average of the power spectrum measured from the randomized maps. Both results are in very good agreement, within $\sim 1\%$, thus validating shot-noise subtraction in our analysis.

\subsection{Unblinding}
\label{sec:unblind}
We validated the pipeline by following all the steps outlined in Section \ref{sec:blinding}. These tests revealed no indication of systematic contamination in our measurements. Subsequently, we unblinded the cosmological parameter contour using real data and confirmed it produced a reasonable PTE with respect to the fiducial model ($>1\%$) using the covariance computed based on fiducial cosmology. 

After unblinding the data, we adjusted the flat-prior range for the linear galaxy bias from [0.8-3.0] to [0.8-3.2]. We made this choice based on preliminary results that indicated the best-fit galaxy bias values of the last three redshift bins were very close to the edge of the prior. This choice doesn't effectively change the results other than making the error bars slightly larger. 

At the start of our work, our approach involved utilizing a binary mask that was generated based on non-zero values observed on a map based on the DES completeness map. However, during the latest stages of our work, we realized that the correct DES mask necessary for the harmonic space analysis should encompass the completeness of each individual pixel. In light of this, we updated the analysis to incorporate the completeness information as the DES mask, as outlined in Sec. \ref{sec:masks_maps}. Consequently, this modification primarily affected the bandpowers at smaller scales, which were excluded from the main analysis due to scale cuts. As a result, the change on the main $S_8$ value was minimal, $\sim 0.1 \sigma$. Nevertheless, we show the results after implementing the updated mask within our pipeline.

At this stage, the parameter inference was done using the covariance computed based on fiducial cosmology. Lastly, we updated the parameter inference by using the covariance matrix computed at the best-fit parameter values. In the subsequent sections, we present our results after accounting for these changes.

\section{RESULTS}
\label{sec:results}

\subsection{Power spectra}
\begin{figure}[ht]
\centering
\includegraphics[scale=0.36]{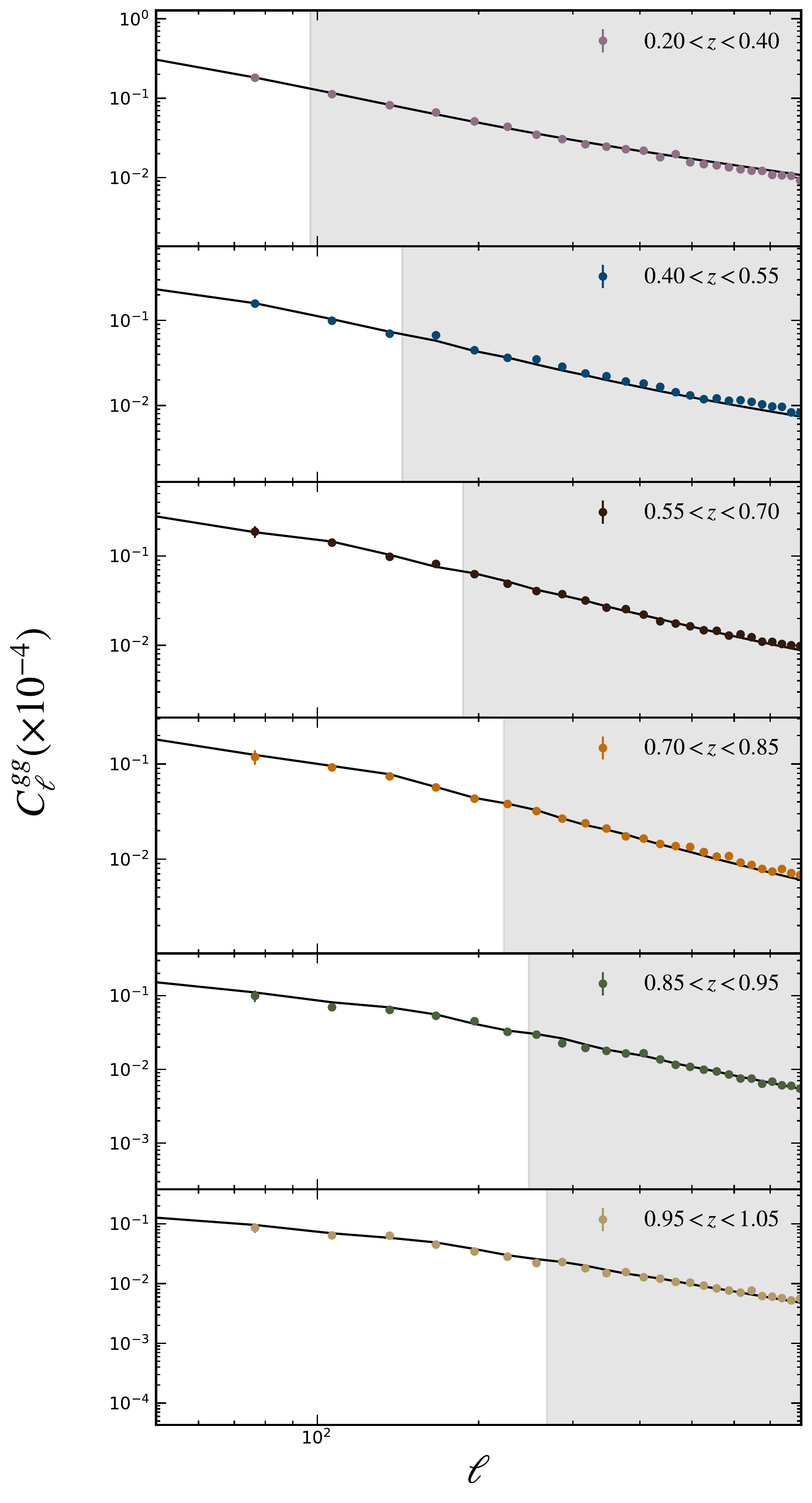}
\includegraphics[scale=0.36]{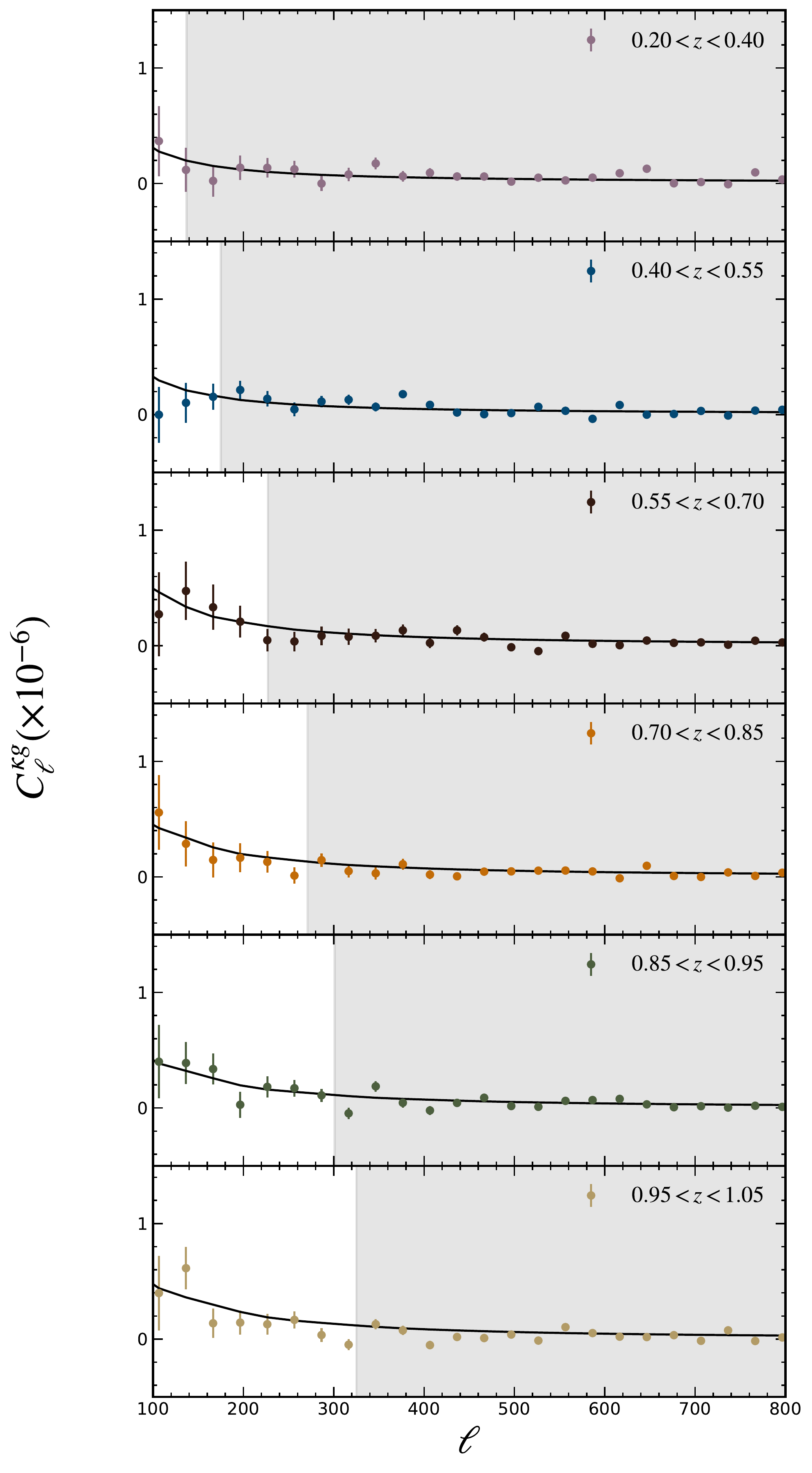}
\caption{Extracted galaxy power spectrum (left panel) and CMB lensing-galaxy power spectrum (right panel) for each redshift bin. The grey-shaded areas highlight the range of multipoles discarded in the baseline analysis. The solid black lines represent theoretical prediction evaluated at the best-fit model developed in Sec. \ref{sec:cosmo_results}.}
\label{fig:cls_realdata}
\end{figure}

Fig. \ref{fig:cls_realdata} shows the measured bandpowers, $C_{\ell}^{gg}$ (left panel) and $C_{\ell}^{\kappa g}$ (right panel), for each tomographic bin of the \maglim sample. The grey shaded region marks the range of multipoles in which the data points were discarded in the main analysis, as discussed in Sec. \ref{sec:method_PS} and \ref{sec:unblind}.

Before presenting the parameter constraints below, we aim to evaluate the total significance of rejecting the null hypothesis of $C_{\ell}^{\kappa g}$, i.e., the significance at which the "no-signal" hypothesis is rejected. This involves computing the significance level of rejecting the null hypothesis as
\begin{equation}
  \chi^2_{\rm null} =  \textbf{D}^T \mathrm{Cov}^{-1}\textbf{D},
    \label{eq:chi_null}
\end{equation}
where the $\textbf{D}$ is the data-vector and $\mathrm{Cov}$ is the corresponding covariance matrix. Here, we consider only $C_{\ell}^{\kappa g}$ as the data vector. Then, we convert the $\chi^2_{\rm null}$ into a probability to exceed (PTE), given the number of degrees of freedom (dof). Finally, we express the PTE in terms of equivalent Gaussian standard deviations as the significance of the rejection of the null hypothesis. We find the significance of the rejection of the null hypothesis is 9.1$\sigma$  when considering the scale cuts of our baseline analysis, and 14.2$\sigma$ when we do not apply any scale cuts ($\ell_{\rm max}$ = $2\times N_{\rm side}$).

\subsection{Linear Galaxy bias}
\label{sec:linear_bias}
The DES Y3 ``3x2pt'' analyses \citep{abbott2018dark}\footnote{The ``3x2pt'' are real-space analyses that include both the auto-correlation of galaxy shear and \maglim galaxies, as well as their cross-correlation.} found that the 2 highest redshift bins show some fluctuations in the measurements that led to a poor fit of the model. Therefore, the main 3x2pt analyses were conducted using only the first four \maglim redshift bins. We can use our $C_{\ell}^{\kappa g}$ measurements to check the consistency of the results, as we expect that CMB lensing cross-correlation would be less (or differently) contaminated by systematic effects that can plague measurements of the galaxy auto-power spectrum. However, since the degeneracy with the galaxy bias prevents the use of $C_{\ell}^{\kappa g}$ and $C_{\ell}^{gg}$ separately to constrain cosmological parameters, we can test the agreement between the two measurements by keeping the underlying cosmology fixed at the fiducial parameters listed in Table \ref{tab:params}, while allowing the galaxy bias and parameters describing the photo-z uncertainties to vary simultaneously. Here, we essentially use the fact that $C_{\ell}^{gg}$ and $C_{\ell}^{\kappa g}$ scale with $b^2$ and $b$, respectively.

\begin{figure}[h!]
    \begin{center}
    \includegraphics[scale=0.5]{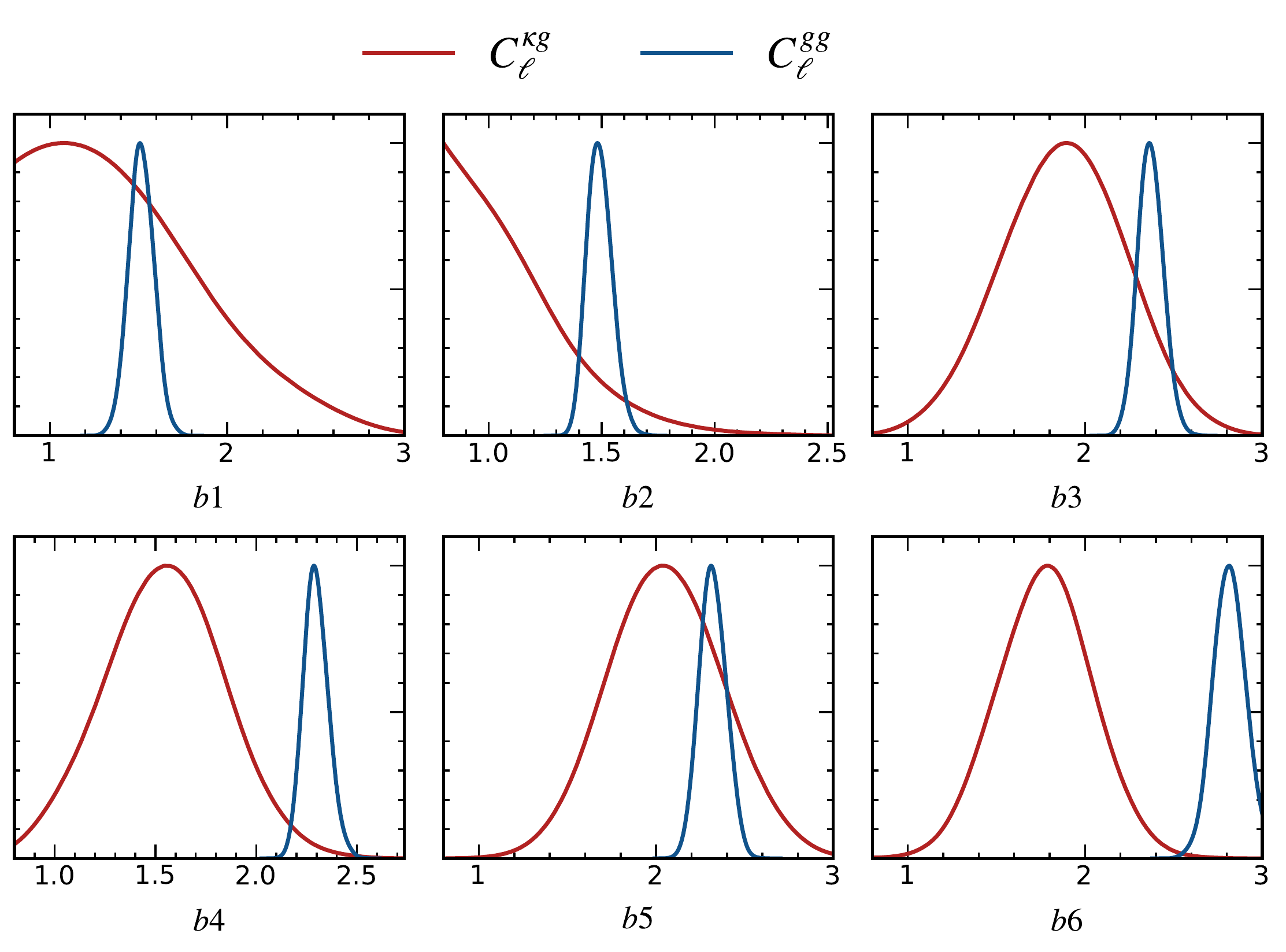}
    \caption{Constraints on linear galaxy bias for the six tomographic bins of the \maglim sample, with cosmological parameters fixed assuming the fiducial cosmology. The blue lines denote the results from galaxy-galaxy and the red lines denote the results from galaxy-CMB lensing power spectrum.}
    \label{fig:bias}
    \end{center}
\end{figure}

Fig. \ref{fig:bias} shows the comparison of the inferred linear galaxy bias from $C_{\ell}^{gg}$ and $C_{\ell}^{\kappa g}$ individually. As expected, the cross-power spectrum has less stringent constraining power than the galaxy power spectrum. When using the $C_{\ell}^{gg}$ alone we find $b_{1}= 1.51\pm 0.07$, $b_{2}= 1.49^{+0.05}_{-0.06}$, $b_{3}= 2.36\pm 0.07$, $b_{4}= 2.29^{+0.05}_{-0.06}$, $b_{5}= 2.32\pm 0.07$, and $b_{6}= 2.80^{+0.08}_{-0.09}$, with the best-fit $\chi^2_{\rm bf}=$ 25.2 for 26 data points corresponding to a PTE of $50.7\%$. When considering the $C_{\ell}^{\kappa g}$ alone we obtain $b_{1}= 1.63{\pm 0.4}$, $b_{2}= 1.20{\pm 0.22}$, $b_{3}= 1.88{\pm 0.35}$, $b_{4}= 1.55{\pm 0.30}$, $b_{5}= 2.05{\pm 0.32}$, $b_{6}= 1.77{\pm 0.27}$, with $\chi^2_{\rm bf}=42.1$ for 31 data-points and PTE = $8.81\%$. 

For all the redshift bins, the galaxy bias from $C_{\ell}^{gg}$ exhibits a preference for a higher galaxy bias compared to the $C_{\ell}^{\kappa g}$. However, given the uncertainties, we observe an acceptable agreement between the two results, except for the highest redshift bin. Other studies indicated potential issues in the measurements or modeling of the two high-redshift \maglim bins \citep{porredon2021dark_maglim, DESXSPT}, which can be connected to this result. However, these analyses rely on the two-point correlation function in real space, and the results largely depend on the assumed fiducial cosmology and the scale cuts applied. 

Despite this result, in Sec. \ref{sec:cosmo_results}, we explore the stability of the cosmological constraints by using (and removing) an individual tomographic bin, finding no evidence of a significant shift. We expect to have a hint of this difference in the future using the DR6 ACT CMB lensing map which will enable us to infer a higher signal-to-noise cross-correlation. 

\subsection{Linear growth of structure}
\label{sec:dg_method}
Given the measured galaxy and galaxy-CMB lensing power spectra, we compute the linear growth factor for the six redshift bins. To this end, we use the $\hat{D}_{G}$ estimator \cite{giannantonio2016cmb}, which allows one to break the degeneracy between the galaxy bias and growth structure under the assumption that both are linear and do not evolve within the selected redshift bin. 

In the $\hat{D}_{G}$ computation, we account for the errors associated with the bandpowers by rewriting Eq. \ref{eq:dg} as a weighted average across multipoles,
\begin{equation}
    \hat{D}_{G} = \frac{\sum_{L}w_{L}\hat{D}_{\rm {G,L}}}{\sum_{L}w_{L}},
\label{eq:dg_data}
\end{equation}
where $\hat{D}_{G}$ for each band power $L$ is given by
 
\begin{equation}
    \hat{D}_{G,L} =  \frac{(C_{L}^{\kappa g})_{\rm obs}}{(\slashed{C}_{L}^{\kappa g})_{\rm th}}\sqrt{\frac{(\slashed{C}_{L}^{gg})_{\rm th}}{(C_{L}^{gg})_{\rm obs}}}.
\label{eq:DGL}
\end{equation}
The weights are expressed in terms of the error of the power spectrum as
\begin{equation}
    w_{L}^{-1} = \hat{D}_{G,L}^2 \bigg[\bigg(\frac{\sigma(C_{L}^{\kappa g})_{\rm obs}}{(C_{L}^{\kappa g})_{\rm obs}}\bigg)^2 + \frac{1}{4} \bigg(\frac{\sigma(C_{L}^{gg})_{\rm obs}}{(C_{L}^{gg})_{\rm obs}}\bigg)^2 \bigg].
\end{equation}

\begin{figure}[ht]
    \centering
    \includegraphics[scale=0.6]{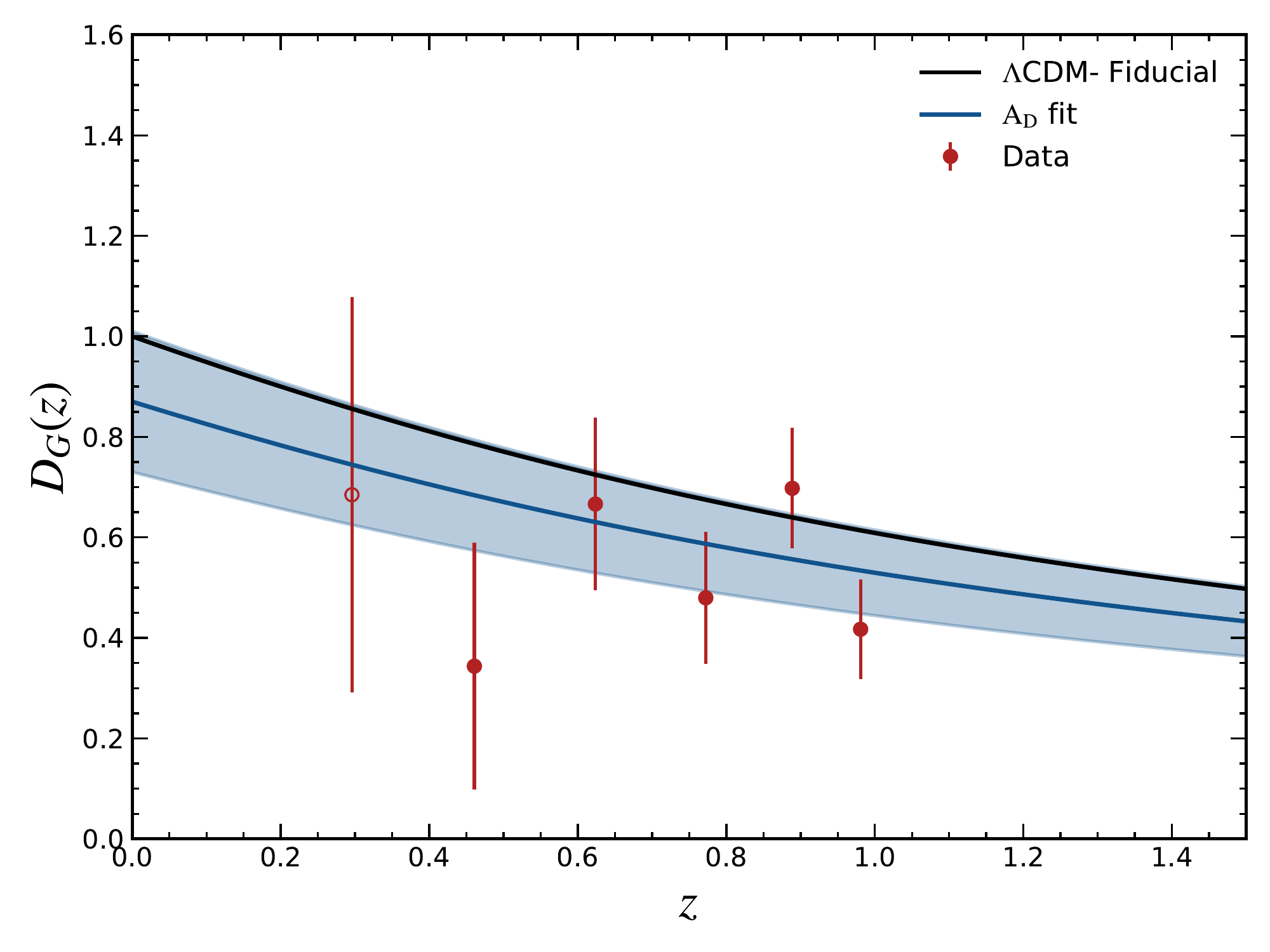}
    \caption{The linear growth factor of each tomographic bin estimated with the $\hat{D}_{\rm G}$ statistic (Eq. \ref{eq:dg_data}) that uses the combination of the galaxy and CMB-lensing-galaxy power spectrum. The red points correspond to the $\hat{D}_{\rm G}$ estimated from the data within the scale range described in Sec \ref{sec:dg_method}. The open point at $z\sim 0.29$ corresponds to the constraint using the weakly non-linear scales described in the text and is not included to fit the growth amplitude $A_{D}$, which is shown by the blue line. The blue band represents the $1\sigma$ confidence interval on the best-fit amplitude $A_{D}$, assuming the fiducial cosmology shown in the solid black line.}
     \label{fig:dg_fromdata}
\end{figure}

The $\hat{D}_{\rm G}$ estimator is computed by taking the average of the bandpowers over a multipole range. To avoid nonlinearities, we set the maximum multipole to $\ell_{\rm max}^{gg}$, as is specified in Table \ref{table:specifics}, while the large-scale cut is set to $\ell_{\rm min} = 100$ due to limitations in the $\kappa$ map. However, applying these scale cuts, the power spectra of the first redshift bin would not lie in a range suitable to compute $D_{G}$. Instead, we consider weakly-nonlinear scales up to $\ell_{\rm max}=137$ for this particular bin, but we exclude this point from the main fit below and use it solely for plotting purposes. We show in Fig. \ref{fig:dg_fromdata} the linear growth factor estimated for each tomographic bin with the corresponding $1\sigma$ error bar. The error bars are estimated from the dispersion of the $\hat{D}_{\rm G}$ computed using the auto- and cross-spectra of the correlated simulations described in Sec. \ref{sec:logmocks}. 

From the growth factor estimated in each redshift bin, we can assess the evolution of growth assuming that the theoretical growth function $D_{G}^{\rm fid}(z)$ scale linearly with a redshift-independent amplitude $A_{D}$. Thus, $A_{D}$ probes the growth function with respect to the fiducial cosmology and can be estimated by minimizing the following $\chi^2$ 
\begin{equation}
    \chi^2 = \sum_{ij}^{6} (\hat{D}_{G}(z_{i}) - A_{D}D_{G}^{\rm fid}(z_{i}))\mathrm{Cov}^{-1}_{ij}(\hat{D}_{G}(z_{j}) - A_{D}D_{G}^{\rm fid}(z_{j})).
    \label{eq:fit_parametrized}
 \end{equation}

We use the MCMC approach to fit $A_{D}$, simultaneously varying both $A_{D}$ and the photo-z uncertainties ($\Delta z$ and $\sigma_{z}$) over the priors specified in Table \ref{tab:params}. In this process, we keep the cosmological parameters fixed at their fiducial values. We set a flat prior for the parameter $A_{D}$, ranging from 0.1 to 3. The covariance is estimated from the $\hat{D}_{G}$ computed using the simulations. We obtain $A_{D}$= 0.87$\pm$ 0.14, when we consider the estimates of the five redshift bins, shown as the filled points in Fig \ref{fig:dg_fromdata}. We show in Fig. \ref{fig:dg_fromdata} the fit of the growth amplitude in its $1\sigma$ confidence interval as the blue line and bands, respectively. The obtained value is within $1\sigma$ of the expected value in the fiducial case, $A_{D} =1$. Also, this result is in agreement with similar analyses using other CMB lensing and galaxy samples spanning different redshift ranges \citep{giannantonio2016cmb,bianchini2018constraining,omori2019dark,marques2020tomographic,desy3_beyondlcdm}.

\subsection{Cosmological Constraints}
\label{sec:cosmo_results}
We next fit cosmological parameters to our measurements. For these fits, we sample the posterior simultaneously varying the galaxy bias, the photo-z uncertainties (i.e. $\Delta_{z}$ and $\sigma_{z}$), and the flat-$\Lambda$\rm{CDM} parameters: $\Omega_{m}$, $\Omega_{b}$, $n_{s}$, $A_{s}$, and $h$. In total, this inference marginalizes over 23 parameters with prior ranges listed in Table \ref{tab:params}. Typically LSS data is especially
sensitive to $\Omega_m$ and to the combination  
\begin{equation}
    S_{8} \equiv \sigma_{8}\sqrt{\frac{\Omega_{m}}{0.3}},
\end{equation}
which we are interested in placing constraints.

\begin{figure}
    \begin{center}
    \includegraphics[scale=1.1]{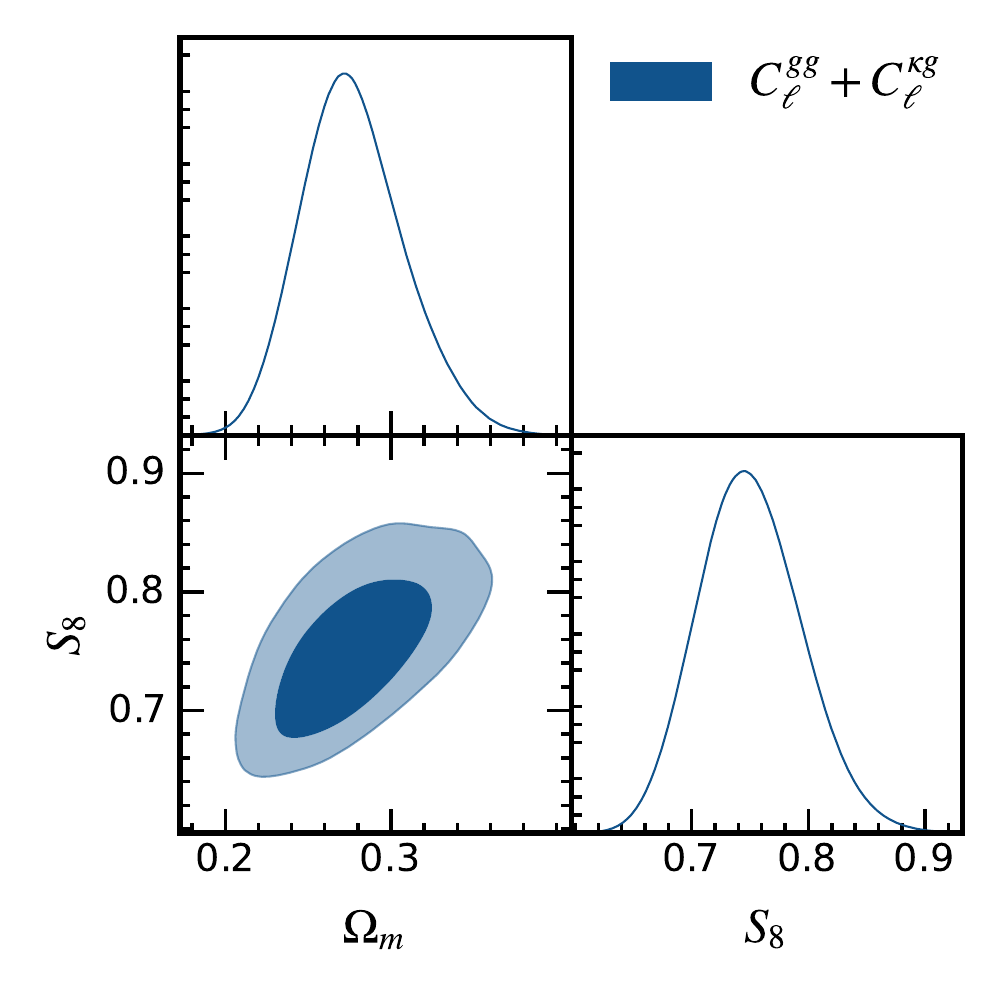}
    \caption{Constraints on $\Omega_m$ and $S_8$ obtained from the joint $C_{\ell}^{gg}$ and $C_{\ell}^{\kappa g}$ of the six tomographic redshift bins of the sample. Both $68\%$ and $95\%$ credible levels are shown.}
    \label{fig:s8_baseline}
    \end{center}
\end{figure}

Figure \ref{fig:s8_baseline} shows the constraints on the $\Omega_{m}$ and $S_8$ from the measured $C_{\ell}^{gg}$ and $C_{\ell}^{\kappa g}$ of the 6 tomographic bins of the \maglim sample. We find that the marginalized 68$\%$ C.L. mean values (best-fit values inside parentheses) are:

\begin{itemize}
\centering
\item[] $\Omega_m = 0.277 ^{+0.029}_{-0.034} (0.26)$;
\item [] $S_{8} = 0.751^{+0.046}_{-0.048} (0.73)$.
\end{itemize}
 
The best-fit model has $\chi^2 = 48.7$ for 57 dof, corresponding to a PTE 77.4$\%$. 

To ensure the consistency of the results, we repeat the analysis considering various combinations of data. 
We perform the parameter constraints considering the following setups: 
 \begin{itemize}
    \item The measurements of each individual photo-z bin. 
    \item The measurements of 5 photo-z bins, excluding one \maglim photo-z bin at a time.
    \item The combination of the 4 lowest redshift bins, i.e., removing the two highest  photo-z bins from the analysis. 
\end{itemize}

\begin{figure}[ht]
    \centering
    \subfloat{{\includegraphics[scale=0.5]{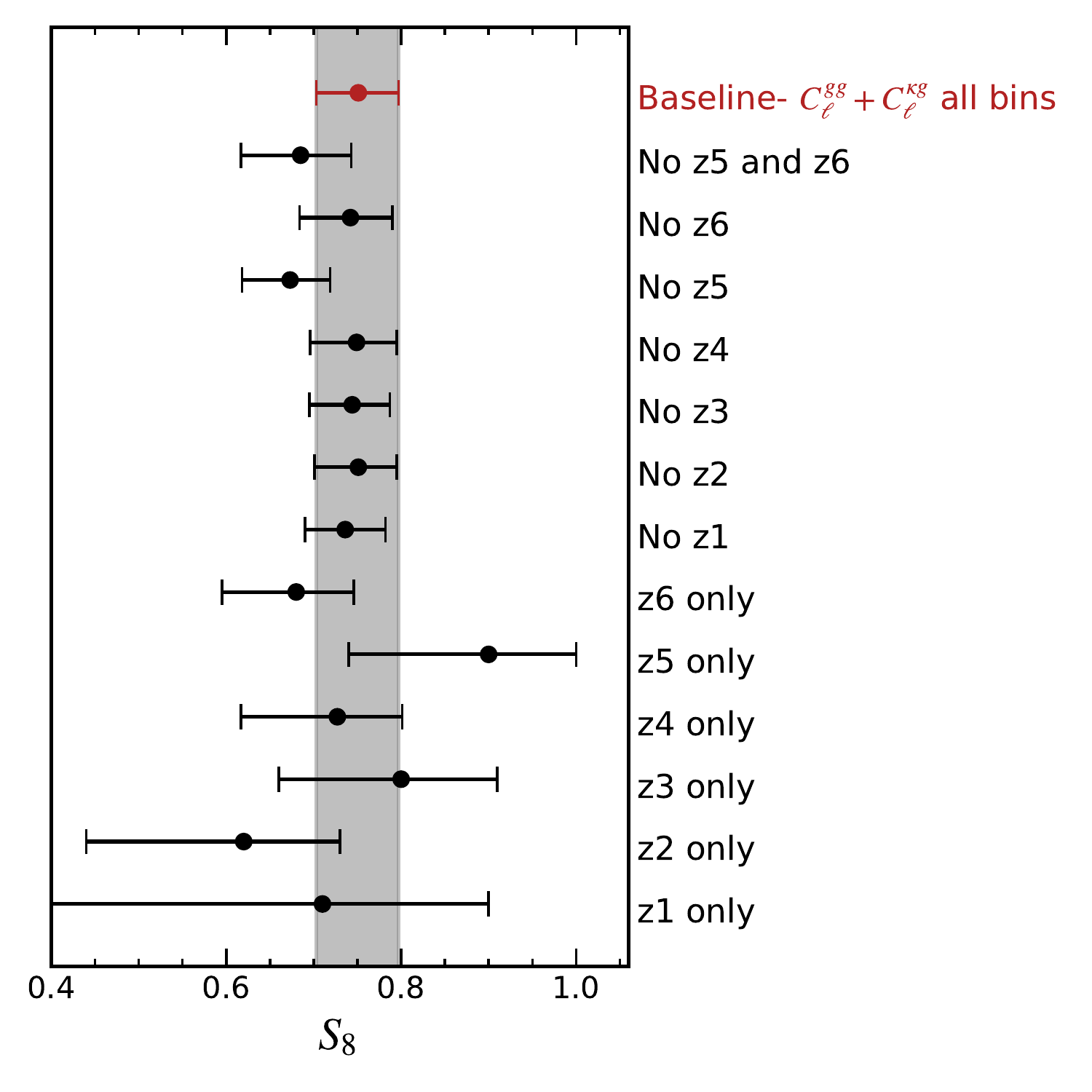} }}%
    \subfloat{{\includegraphics[scale=0.5]{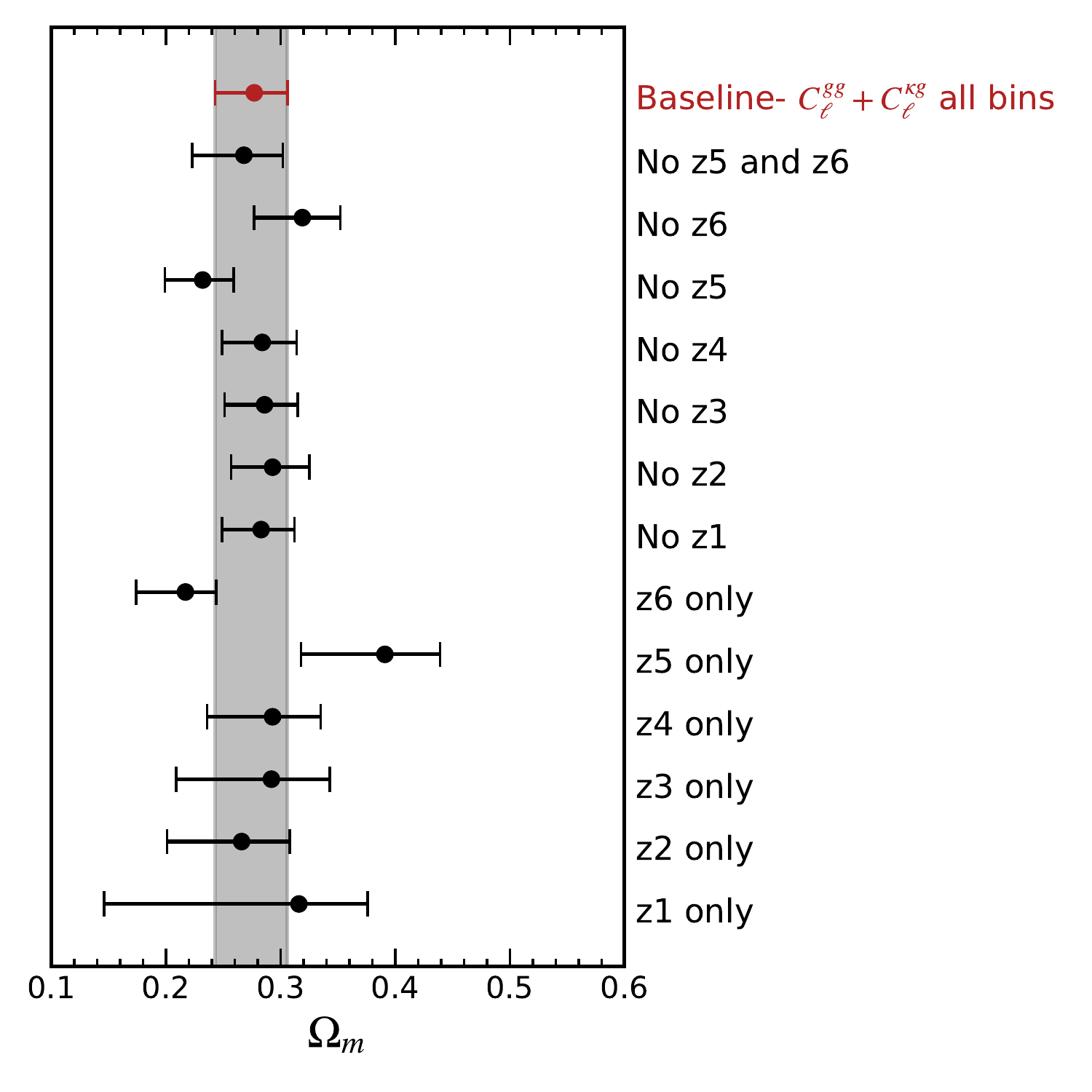} }}%
    \caption{Summary of the constraints on $S_8= \sigma_8 (\Omega_m /0.3)^{0.5}$ (left panel) and $\Omega_{m}$ (right panel)  and their robustness against different setups in the data combination (see Section \ref{sec:cosmo_results} for more details). The red dot denotes the baseline result.}%
    \label{fig:summary_S8s}%
\end{figure}

Fig.\ref{fig:summary_S8s} shows the $1\sigma$ uncertainty on $S_{8}$ and $\Omega_{m}$ for the different scenarios described above. As a comparison, we also show the result of the baseline analysis presented highlighted in red. The largest parameter shift towards lower $S_8$ is found when we use only measurements from the second photo-z bin to establish constraints on the cosmological parameters; on the other hand, using the fifth photo-z bin shifts the $S_{8}$ to higher values than the baseline result. However, these constraints have visibly larger uncertainties in both cases and the shifts are less than $1\sigma$ different from the baseline result. It is worth noticing that in Fig. \ref{fig:dg_fromdata} is possible to see a similar behavior, in which the growth rate of the second (and of the fifth) photo-z bin is relatively lower (higher) than the other points. Given Fig. \ref{fig:summary_S8s}, this is not a surprising result since the $D_{G}$ scales with $S_{8}$ and $\Omega_{m}$ as well. In addition, the results are stable when removing one of the photo-z bins at a time. Therefore, the parameter constraints for all these setups are consistent with the baseline result indicating an internal consistency of the baseline data choice. 

Recent studies using the two-point information of the DES Y3 \maglim sample also add constraints on $S_8$, including the galaxy clustering and galaxy-galaxy lensing \citep{porredon2021dark_maglim} ($S_8 = 0.778^{+0.037}_{-0.031}$) and the 3x2pt \citep{descollaboration2021dark} ($S_8 = 0.776\pm 0.017$). In \cite{DESXSPT}, $S_8$ is derived from the cross-correlation between \maglim galaxies and DES Y3 shear with the CMB lensing from \textit{Planck} and SPT ($S_8 = 0.736^{+0.032}_{-0.028}$), while \citep{combined_des_spt} consider the combination of the 2-point correlation function between SPT CMB-lensing, galaxy positions, and galaxy lensing ($S_8 = 0.792\pm 0.012$). Despite the fact that these analyses use different combinations of data and employ real-space correlation functions, which makes it difficult to compare their analysis choices with ours in harmonic space, our results are statistically consistent with all of them.
  
\begin{figure}[ht]
    \centering
    \includegraphics[scale=0.7]{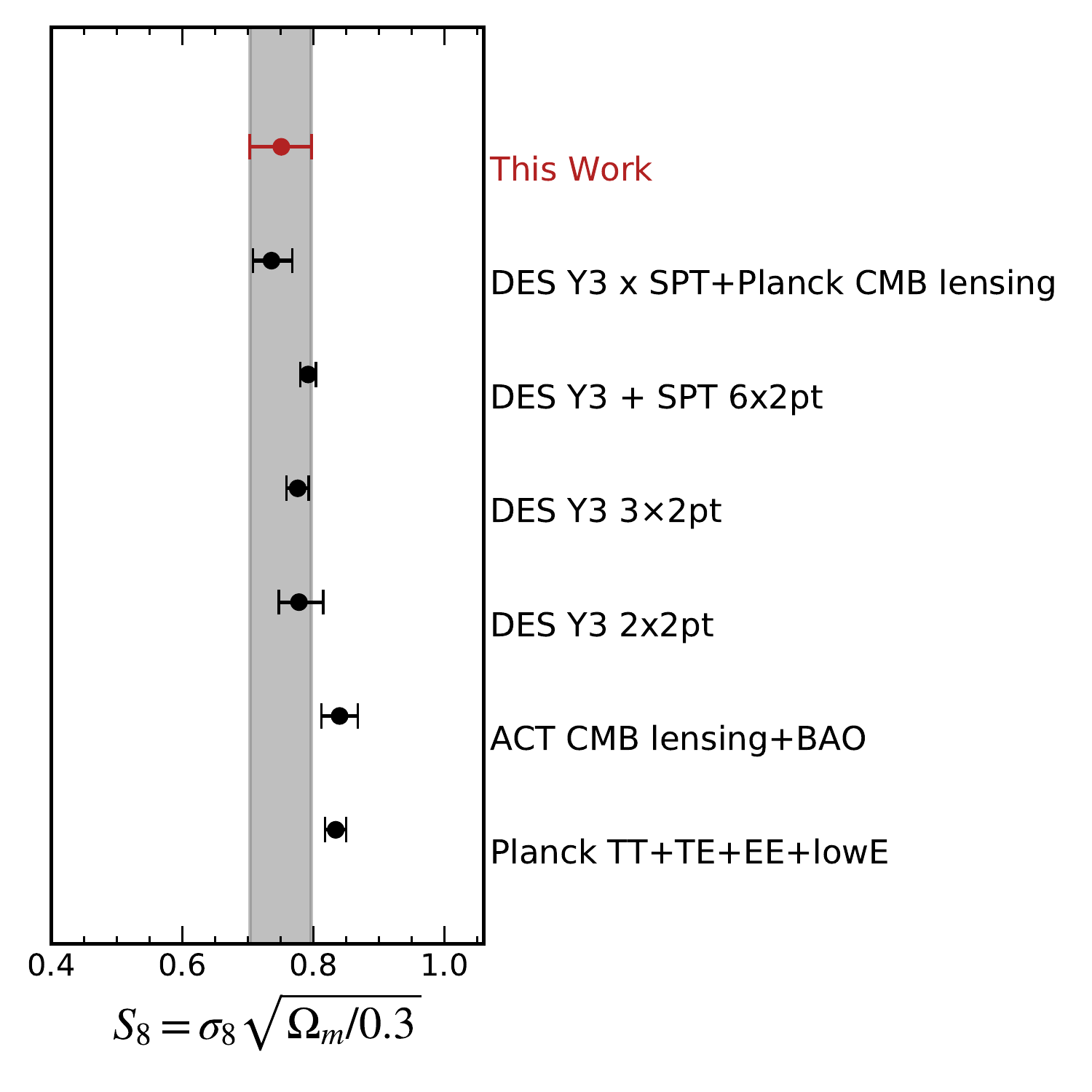}
    \caption{Comparison of measurement of $S_8$ from our analysis with other measurements using DES Y3 data \citep{DESXSPT,combined_des_spt,descollaboration2021dark,porredon2021dark_maglim}, ACT CMB lensing + BAO  \citep{dr6atacama}, and primordial CMB from \textit{Planck} \citep{planckparams}, though note that the DES points here include marginalization over the sum of neutrino masses.}
    \label{fig:literature_S8}
\end{figure}

Other studies that rely on late-time data also suggest a lower $S_8$ value compared to what is inferred from CMB data within the context of $\Lambda$CDM. In particular, the analysis of the TT-TE-EE- and low-E polarization of the \textit{Planck} satellite (\textit{Planck} TT+TE+EE+lowE) found $S_8=0.834 \pm 0.016$ \citep{planckparams}, which is $\sim 1.7\sigma$ different from our results when adding the statistical uncertainties in quadrature. Interestingly, analysis of the CMB lensing from \textit{Planck} \citep{planckcmblens} and, more recently, from ACT \citep{qu2023atacama,dr6atacama}, also reveals a $S_{8}$ value consistent with the CMB constraints, suggesting that tracers at higher redshifts and probing larger scales tend to prefer higher $S_8$ than the probes at lower redshifts. In Fig. \ref{fig:literature_S8} we summarize the comparison of our 1D marginalised $S_8$ parameter against these measurements. 

Our results are also in agreement with other CMB-lensing cross-correlation studies, such as with the unWISE ($S_8= 0.784\pm 0.015$) \citep{krolewski2021cosmological}, DESI ($S_8= 0.73\pm 0.03$) \citep{white2022cosmological}, and KiDS ($S_8 = 0.64\pm 0.08$) \citep{robertson2021strong} data. The results are also consistent with \cite{garcia2021growth}, who reported a low $S_8$ value ($S_8 = 0.7781 \pm 0.0094$) in comparison to CMB constraints through the cross-correlation of \textit{Planck} CMB lensing with various galaxy surveys. However, their results exhibited higher $S_8$ when excluding shear measurements ($0.825 \pm 0.023$). 

It remains unclear whether the observed discrepancy in the $S_8$ values measured from the CMB and those derived from the late universe is due to a physical reason, an unaccounted for systematics, or due to a statistical fluctuation. It is thus important to emphasize that our analysis relies on a few assumptions. We assume validity of the model in the linear regime so that the galaxy overdensity is connected to the matter overdensity by a linear, deterministic, and scale-independent galaxy bias. However, even applying conservative scale cuts to ensure linearity, this relationship may no longer be accurate if there is a stochastic component in the galaxy density, which could be caused by various observational and astrophysical factors, such as discrete sampling and physical effects on galaxy formation other than those from the local density field \citep{tegmark1998time}. To account for these potential effects, it would be necessary to increase the complexity of the model. Nevertheless, due to the uncertainties in our measurements, it is unlikely that we would be able to draw a conclusive result even with an improved model. Therefore, we leave this for future work.

\section{SUMMARY}
\label{sec:conclusions} 

We have presented the first analysis in harmonic space of the cross-correlation between the galaxy density fluctuations from DES Y3 \maglim galaxies and CMB lensing cross-correlation. We use the CMB convergence map from the fourth ACT data release. This measurement, when combined with the galaxy power spectrum, helps to break the degeneracy between the galaxy bias and cosmological parameters. We have derived cosmological constraints by considering a tomographic approach, with six redshift bins spanning from $z= 0.20$ to $1.05$.  

Our pipeline has been extensively validated and we carried out several systematic checks (Sec.\ref{sec:test_logmocks} and \ref{sec:sys}), in which we found no significant evidence of unaccounted contamination in our measurements within the range of scales we use. Initially, we performed the analysis under a blinding procedure described in \ref{sec:blinding}. After applying very conservative cuts to ensure our data lies in the linear regime, we rejected the null hypothesis of no correlation at 9.1$\sigma$. The bandpowers measurements are displayed in Fig. \ref{fig:cls_realdata}.

We use our measurements to constrain cosmology under different scenarios. First, we fixed the cosmology at the fiducial values and checked the consistency of the linear galaxy bias constrained from the $C_{\ell}^{\kappa g}$ and from the $C_{\ell}^{gg}$ data. The main result is displayed in Fig.~\ref{fig:bias} and indicates a acceptable concordance in all redshift bins, with the exception of the highest redshift bin. In this particular bin, we observe an indication of a potential discrepancy between the linear galaxy bias derived from our two measurements. Nevertheless, when removing this particular redshift bin to place constraints on $S_8$, we did not observe any significant impact on the results.

Next, we investigated the linear growth rate of the structures at different redshifts by employing the galaxy bias-independent $D_{G}$ estimator defined in Sec. \ref{sec:dg_method}. The main result is shown in Fig. \ref{fig:dg_fromdata}. Given the growth measurements at each redshift, we constrain the parameter $A_D$, which represents the total amplitude of the linear growth with respect to the fiducial cosmology. We find $A_D = 0.87 \pm 0.14$, consistent with the expected value of the fiducial model, ($A_D =1$). Measuring $D_G$ tomographically is a possible way to investigate whether a preferred lower or higher $S_8$ value exists within a specific redshift range. This approach is motivated by the fact that certain analyses with late-time data suggest lower values of the clustering amplitude compared to the constraints imposed by primordial CMB. Due to the data limitations and the conservative scale cuts applied, we are not able to place constraints strong enough to further examine if there is a preferred trend of the $D_G$ evolution. This approach would significantly benefit from the use of a convergence CMB lensing map with improved overlap and noise levels, such as the latest ACT data release 6 \citep{qu2023atacama, dr6atacama} that almost entirely overlaps with the DES region \citep{darwishcrossdes}.

Finally, we vary the cosmological and astrophysical parameters to place constraints on $\Omega_m$ and $S_8$. In $\Lambda$CDM, we find at $68\%$ C.L  $\Omega_m = 0.277 ^{+0.029}_{-0.034}$, and $S_{8} = 0.751^{+0.046}_{-0.048}$. Our main result is shown in Figure \ref{fig:s8_baseline}. We find that our $S_8$ constraints are slightly lower with respect to \textit{Planck} TT+TE+EE+lowE at the $\sim 1.7 \sigma$ level when adding the statistical uncertainties in quadrature. Our result is also consistent with other studies \citep{descollaboration2021dark,DESXSPT,porredon2021dark_maglim,krolewski2021cosmological,white2022cosmological,robertson2021strong}. We perform a number of internal consistency tests to assess the stability of the main result. We conducted several internal consistency tests to evaluate the robustness of our main result. A summary of these tests is presented in Fig. \ref{fig:summary_S8s}.

The current discrepancy between the inferred value of $S_{8}$ from the CMB and from some late Universe observations emphasizes the need for testing the $\Lambda$CDM model and systematic effects through various methods, datasets, and pipelines. Our study demonstrates the ability to constrain cosmology using the combination of the galaxy power spectrum and the CMB lensing- galaxy cross-power spectrum. In a companion paper \citep{shabbirDES&ACT}, we explore the parameter constraint using the cross-correlation between ACT DR4 CMB lensing and DES Y3 shear data, which is also sensitive to the amplitude of large-scale structure parameter $\sigma_8$. Furthermore, the next data releases of galaxy surveys such as the DES, HSC, DESI, and new CMB lensing maps such as from AdvACT and SPT-3G will tighten the parameter constraints and help us to understand the history of the cosmic structure growth. This will be especially valuable if one can demonstrate that the systematic effects are well controlled, even with increased precision. In the future, a comprehensive cross-correlation analysis will be even more stringent using data from upcoming surveys such as the Vera Rubin Observatory Legacy Survey, Euclid mission \footnote{\url{https://www.euclid-ec.org/}}, the Nancy G. Roman Space Telescope\footnote{\url{https://roman.gsfc.nasa.gov/}}, Simons Observatory\footnote{\url{https://simonsobservatory.org/}}, and CMB-S4\footnote{\url{https://cmb-s4.org/}}. However, accurate theoretical modeling is crucial for achieving this goal. Although we have adopted conservative scale cuts and do not extend the analysis to include mildly-small scales, a comprehensive understanding of effects significant in this regime (e.g., nonlinear galaxy bias, baryonic effects) will be essential to performing precision cosmology and is something to be addressed in a higher signal-to-noise measurement using ACT DR6 cross-correlation studies \cite{darwishcrossdes}.

\appendix

\acknowledgments

Support for ACT was through the U.S.~National Science Foundation through awards AST-0408698, AST-0965625, and AST-1440226 for the ACT project, as well as awards PHY-0355328, PHY-0855887 and PHY-1214379. Funding was also provided by Princeton University, the University of Pennsylvania, and a Canada Foundation for Innovation (CFI) award to UBC. ACT operated in the Parque Astron\'omico Atacama in northern Chile under the auspices of the Agencia Nacional de Investigaci\'on y Desarrollo (ANID). The development of multichroic detectors and lenses was supported by NASA grants NNX13AE56G and NNX14AB58G. Detector research at NIST was supported by the NIST Innovations in Measurement Science program.  Computing for ACT was performed using the Princeton Research Computing resources at Princeton University, the National Energy Research Scientific Computing Center (NERSC), and the Niagara supercomputer at the SciNet HPC Consortium.

Funding for the DES Projects has been provided by the U.S. Department of Energy, the U.S. National Science Foundation, the Ministry of Science and Education of Spain, 
the Science and Technology Facilities Council of the United Kingdom, the Higher Education Funding Council for England, the National Center for Supercomputing 
Applications at the University of Illinois at Urbana-Champaign, the Kavli Institute of Cosmological Physics at the University of Chicago, 
the Center for Cosmology and Astro-Particle Physics at the Ohio State University,
the Mitchell Institute for Fundamental Physics and Astronomy at Texas A\&M University, Financiadora de Estudos e Projetos, 
Funda{\c c}{\~a}o Carlos Chagas Filho de Amparo {\`a} Pesquisa do Estado do Rio de Janeiro, Conselho Nacional de Desenvolvimento Cient{\'i}fico e Tecnol{\'o}gico and 
the Minist{\'e}rio da Ci{\^e}ncia, Tecnologia e Inova{\c c}{\~a}o, the Deutsche Forschungsgemeinschaft and the Collaborating Institutions in the Dark Energy Survey. 

The Collaborating Institutions are Argonne National Laboratory, the University of California at Santa Cruz, the University of Cambridge, Centro de Investigaciones Energ{\'e}ticas, 
Medioambientales y Tecnol{\'o}gicas-Madrid, the University of Chicago, University College London, the DES-Brazil Consortium, the University of Edinburgh, 
the Eidgen{\"o}ssische Technische Hochschule (ETH) Z{\"u}rich, 
Fermi National Accelerator Laboratory, the University of Illinois at Urbana-Champaign, the Institut de Ci{\`e}ncies de l'Espai (IEEC/CSIC), 
the Institut de F{\'i}sica d'Altes Energies, Lawrence Berkeley National Laboratory, the Ludwig-Maximilians Universit{\"a}t M{\"u}nchen and the associated Excellence Cluster Universe, 
the University of Michigan, NSF's NOIRLab, the University of Nottingham, The Ohio State University, the University of Pennsylvania, the University of Portsmouth, 
SLAC National Accelerator Laboratory, Stanford University, the University of Sussex, Texas A\&M University, and the OzDES Membership Consortium.

Based in part on observations at Cerro Tololo Inter-American Observatory at NSF's NOIRLab (NOIRLab Prop. ID 2012B-0001; PI: J. Frieman), which is managed by the Association of Universities for Research in Astronomy (AURA) under a cooperative agreement with the National Science Foundation.

The DES data management system is supported by the National Science Foundation under Grant Numbers AST-1138766 and AST-1536171.
The DES participants from Spanish institutions are partially supported by MICINN under grants ESP2017-89838, PGC2018-094773, PGC2018-102021, SEV-2016-0588, SEV-2016-0597, and MDM-2015-0509, some of which include ERDF funds from the European Union. IFAE is partially funded by the CERCA program of the Generalitat de Catalunya.
Research leading to these results has received funding from the European Research
Council under the European Union's Seventh Framework Program (FP7/2007-2013) including ERC grant agreements 240672, 291329, and 306478.
We  acknowledge support from the Brazilian Instituto Nacional de Ci\^encia
e Tecnologia (INCT) do e-Universo (CNPq grant 465376/2014-2).

This manuscript has been authored by Fermi Research Alliance, LLC under Contract No. DE-AC02-07CH11359 with the U.S. Department of Energy, Office of Science, Office of High Energy Physics.

GM and KMH acknowledge support from the National Science Foundation award 1815887 and the FSU College of Arts and Sciences Dean’s Postdoctoral Scholar Fellows program. MM acknowledges support from NASA grant 21-ATP21-0145. OD acknowledges support from SNSF Eccellenza Pro- fessorial Fellowship (No. 186879). SS acknowledges support from the Beus Center for Cosmic Foundations. IH acknowledges support from the European Research Council (ERC) under the European Union's Horizon 2020 research and innovation programme (Grant agreement No. 849169). EC acknowledges support from the European Research Council (ERC) under the European Union’s Horizon 2020 research and innovation programme (Grant agreement No. 849169). JCH acknowledges support from NSF grant AST-2108536, NASA grants 21-ATP21-0129 and 22-ADAP22-0145, DOE grant DE-SC00233966, the Sloan Foundation, and the Simons Foundation. KM acknowledges support from the National Research Foundation of South Africa. CS acknowledges support from the Agencia Nacional de Investigaci\'on y Desarrollo (ANID) through FONDECYT grant no.\ 11191125 and BASAL project FB210003.

We acknowledge the use of many public Python packages not cited along the main text: Numpy \citep{oliphant2015guide}, Astropy\footnote{\url{http://www.astropy.org}} a community-developed core Python package for Astronomy~\citep{astropy:2013, astropy:2018}, Matplotlib~\citep{hunter2007matplotlib}, IPython~\citep{perez2007ipython} and Scipy~\citep{jones2001scipy}.

\bibliographystyle{JHEP}
\bibliography{main.bib}

\end{document}